\documentclass[11pt,preprint]{aastex}
\usepackage{graphicx}
\usepackage{amsmath,amssymb} 
\usepackage{subfigure}

\begin{document}

\newcommand{\xhat}{\hat{\boldsymbol x}}
\newcommand{\yhat}{\hat{\boldsymbol y}}
\newcommand{\zhat}{\hat{\boldsymbol z}}
\newcommand{\rhat}{\hat{\boldsymbol r}}
\newcommand{\cphat}{\hat{\boldsymbol \varpi}}
\newcommand{\that}{\hat{\boldsymbol \theta}}
\newcommand{\phat}{\hat{\boldsymbol \phi}}
\newcommand{\bfnabla}{\boldsymbol \nabla}
\newcommand{\bfv}{\boldsymbol v}
\newcommand{\bfxi}{\boldsymbol \xi}
\newcommand{\bfg}{\boldsymbol g}
\newcommand{\ba}{\begin{eqnarray}}
\newcommand{\ea}{\end{eqnarray}}
\newcommand{\eh}{\hat{\boldsymbol e}}
\newcommand{\ov}{\overline}
\newcommand{\ti}{\tilde}
\newcommand{\sgn}{\mathrm{sgn}}
\newcommand{\tk}{\tilde{k}}
\newcommand{\tkp}{\tilde{k}_+}
\newcommand{\tkm}{\tilde{k}_-}
\newcommand{\kp}{\bar{k}_+}
\newcommand{\km}{\bar{k}_-}
\newcommand{\vp}{\varphi}
\newcommand{\eps}{\epsilon}
\newcommand{\cp}{\varpi}
\newcommand{\gb}{\bar{g}}
\newcommand{\bo}{\bar{\omega}}
\newcommand{\bM}{\bar{M}}
\newcommand{\bs}{\bar{\sigma}}
\newcommand{\HF}{{_1}F_1}

\title{Supersonic Shear Instabilities in Astrophysical Boundary Layers}

\author{Mikhail A. Belyaev\altaffilmark{1} \& 
Roman R. Rafikov\altaffilmark{1,2}}
\altaffiltext{1}{Department of Astrophysical Sciences, 
Princeton University, Ivy Lane, Princeton, NJ 08540; 
rrr@astro.princeton.edu}
\altaffiltext{2}{Sloan Fellow}

\begin{abstract}
Disk accretion onto weakly magnetized astrophysical objects 
often proceeds via a boundary layer that forms near the 
object's surface, in which the rotation speed of the 
accreted gas changes rapidly. Here we study the initial 
stages of formation for such a boundary layer around a 
white dwarf or a young star by examining the hydrodynamical 
shear instabilities that may initiate mixing and momentum 
transport between the two fluids of different densities moving
supersonically with respect to each other. We find that 
an initially laminar boundary layer is unstable to two different 
kinds of instabilities. One is an instability of a supersonic 
vortex sheet (implying a discontinuous initial profile of the 
angular speed of the gas) in the presence of gravity, which
we find to have a growth rate of order (but less than) the 
orbital frequency. The other is a sonic instability of a 
finite width, supersonic shear layer, which is similar to the 
Papaloizou-Pringle instability. It has a growth rate proportional 
to the shear inside the transition layer, which is of order 
the orbital frequency times the ratio of stellar radius to 
the boundary layer thickness. For a boundary layer that
is thin compared to the radius of the star, the shear rate is much
larger than the orbital frequency. Thus, we conclude that sonic
instabilities play a dominant role in the initial stages of
nonmagnetic boundary layer formation and give rise to very 
fast mixing between disk gas and stellar fluid in the 
supersonic regime.
\end{abstract}

\keywords{accretion, accretion disks -- hydrodynamics --- waves -- instabilities}

%%%%%%%%%%%%%%%%%%%%%%%%%%%%%%%

\section{Introduction}

Accretion onto astrophysical objects possessing a material surface  
(as opposed to accretion onto black holes) always involves a non-trivial
interaction of the incoming gas with the object's outer layers. 
Examples of such objects include white dwarfs in cataclysmic 
variables (CVs), young stars gaining material from a protoplanetary
disk, and accreting neutron stars. In all these systems, one must 
understand how the incoming gas shares its angular momentum with
the accreting object and how it mixes with the previously accreted 
material. 

If the magnetic field of the central object is strong enough, it can
disrupt the disk at some distance above the surface. Inside this region, 
accretion proceeds along field lines
\citep{GhoshLamb,Koldobaetal}. The
critical value of the magnetic field at the surface of the central object
for magnetic disruption is given by
\ba
B_{*,\text{crit}} = 5 \times 10^4 \left(\frac{\beta}{.5} \right)^{-7/4}
\left(\frac{M_*}{.6 M_\odot}\right)^{1/4} \left(\frac{\dot{M}}{10^{-8}
  M_\odot \ \text{yr}^{-1}}\right)^{1/2}\left(\frac{R_*}{9 \times 10^3
  \ \text{km}}\right)^{-5/4} \text{G}.
\label{Bdisrupt}
\ea
Here, we have used parameters for the mass and radius that are typical
for CVs in outburst \citep{Bergeronetal,Hachisu,Patterson}. The
parameter $\beta$ is a dimensionless factor of order unity that depends on
the model adopted for the disruption of the disk by the stellar
magnetic field \citep{GhoshLamb}.

For $B_* < B_{*,\text{crit}}$, the accretion disk will
extend all the way to the surface of the star, for which there 
is good observational evidence in neutron star low mass X-ray binaries
\citep{GilfanovRevnivtsevMolkov} and in dwarf nova systems in outburst
\citep{WheatleyMaucheMattei}. Since the rotation rate
of the star, $\Omega_*$, is slower than the Keplerian rotation rate at
the stellar surface, $\Omega_K(R_*)$, a boundary layer (BL) will exist
inside of which $d \Omega / d R > 0$ and the rotation profile of
the star attaches smoothly to that of the disk. If the accretion
rate is high enough, it is also possible for material to spread
meridionally to high latitudes forming a spreading layer (SL)
\citep{InogamovSunyaev, PiroBildsten}. For both a BL and a SL, the intense 
energy release in a localized region near the stellar surface leads to
easily observable signatures such as hard spectral components, 
variability of emission, and so on.

A considerable amount of effort has been previously devoted to 
understanding the structure of well-developed, steady-state BLs,
which have been evolving for long enough to establish a smooth
rotation profile in their interior. An outstanding question in the
study of steady-state BLs is 
identifying the mechanism of angular momentum transport in 
the layer. Several potential mechanisms have been explored over the 
years, among them shear instabilities \citep{KippenhahnThomas}, 
baroclinic instabilities \citep{Fujimoto, Hanawa}, and the
Tayler-Spruit dynamo
\citep{Spruit, PiroBildsten1}. Despite this effort, no clear answer exists at
present regarding the nature of the angular momentum transport and 
mixing in well-developed BLs.

\subsection{Initiation of Boundary Layers}

An interesting aspect of the BL problem that has received less
attention in the past is the issue of BL {\it initiation}, i.e. 
the initial stage of BL formation which must occur when the accreted
material first touches the surface of the star. In this case the physical 
setup is going to be quite different from the steady-state BL 
primarily because of the much larger velocity shear, justifying 
the study of BL initiation in the limit of an
essentially discontinuous rotation profile at the interface between the 
star and accreting material. 

Even though the initiation stage is just a transient phase in the BL 
evolution, one still needs to understand it to get a full picture of
the BL phenomenon. Also, one need not think that the BL initiation 
is a unique event for every accreting object --- it may, in fact, be 
repetitive. The most common situation in which this BL initiation is 
recurrent involves the dramatic increase of the mass accretion 
rate (due to some sort of outburst triggered by an instability) 
through the disk which is magnetically disrupted outside the star 
under normal circumstances. According to Equation (\ref{Bdisrupt}), a
sudden increase of $\dot M$ by several orders of magnitude as typical 
for some accreting systems (e.g. dwarf novae, FU Ori outbursts, etc.) 
can rapidly compress the stellar magnetosphere to the point at which 
disk material starts touching the stellar surface and a BL starts 
to form.

There is good observational evidence for this kind of recurrent 
behavior. For instance, \citet{LivioPringle} have
argued that the observed lag in the rise time of the UV emission
relative to the optical in a CV system transitioning to outburst can
be explained if the disk is magnetically disrupted in quiescence but
not in outburst. Thus, the hottest, innermost part of the disk is
evacuated in quiescence, and UV radiation is not emitted immediately in
the transition to outburst, since the empty region requires time to
become filled. 

FU Ori stars are another type of system in which the
disk can extend all the way to the stellar surface, due to the
high accretion rate - $\dot{M} \sim 10^{-4} M_\sun
\text{yr}^{-1}$ \citep{KHH}. 
In these systems, the BL can puff up to become
of order the stellar radius and the distinction between the boundary
layer, the disk, and the star becomes blurred \citep{PNFuOri}.

The goal of our present work is to make a first step towards 
understanding the {\it formation} of the BL. As in the case of a 
well-developed BL, the major issue for this initial stage lies 
in details of the angular momentum transport and mixing. However,
because of the enormous shear present at the star-disk interface 
at this stage, it is highly likely that purely hydrodynamical shear
instabilities would dominate the transport of mass and momentum
rather than anything else. 

For that reason, in this work we primarily focus on exploring the 
properties of various shear-driven instabilities under the conditions 
typical at the BL formation stage. {First, we seek to identify a
particular variety of the shear instability that most efficiently 
initiates mixing between the two fluids, i.e. has the fastest growth 
rate. Second, we examine the conditions needed for its operation, 
such as the density contrast between fluids, initial velocity 
profile, etc.}

There are several physical 
ingredients which can potentially affect operation of shear
instabilities: stratification, rotation, magnetic fields, turbulence, 
radiative transfer, and the supersonic nature of the flow. The latter 
aspect of the problem is very important and is inevitable during
the BL initiation phase, when disk material rotating at the 
Keplerian speed comes into contact with the more slowly spinning stellar 
surface. The differential azimuthal velocity between the two 
interacting flows is then bound to be a significant fraction of the 
Keplerian velocity at $R_*$ and should highly exceed the  
sound speed both in the disk and in the outer layers of the star.

\subsection{Shear Instabilities in Compressible Fluids}

The study of shear instabilities in compressible fluids is of
fundamental physical significance and has received much attention in the
past. \citet{Landau}, \citet{Hatanaka}, \citet{Pai}, \citet{MilesKH},
and \citet{Gerwin} have all studied the problem in the vortex sheet 
approximation, where one half plane of compressible
fluid moves at constant velocity over another. When the two fluids
move at a low Mach
number relative to each other, one finds that infinitesimal
perturbations are governed by the
classical Kelvin-Helmholtz (KH) dispersion relation for two incompressible
fluids. However, \citet{MilesKH} showed that above a critical Mach
number, the vortex sheet
becomes marginally stable to two-dimensional disturbances along the
direction of the flow. This is a surprising result considering that
one might have expected increasing the shear to lead to
an increased growth rate for the instability rather than stabilization. 
Understanding the implications of this result for the momentum and
mass transport in the astrophysical BLs is one of the goals of our
study.

Later, \citet{Blumenetal}, \citet{Ray}, \citet{ChoudhuryLovelace} and
\citet{Glatzel} studied
the stability of a fluid with a continuous and monotonically varying
velocity profile. They found that unlike the vortex sheet, continuous
velocity profiles with a supersonic velocity difference across them
were unstable even at high Mach number. \citet{Glatzel} showed that
the instability was similar to the
Papaloizou-Pringle instability which operates in hydrodynamical disks with
radial boundaries
\citep{PP}. Thus, for high Mach number flow in a compressible fluid, a
finite thickness velocity
profile exhibits fundamentally different behavior from the vortex
sheet, since it can support unstable modes. Moreover, the growth rate
of the unstable modes scales as $\propto 1/L$, where $L$ is the
thickness of the shear layer. Thus, the thinner the shear layer, the
faster the instability operates! This is exactly the opposite of what
one might have expected from the vortex sheet stability criterion and
we will provide an explanation for this in \S \ref{finite_width_SL}.

Previous studies of possible instabilities inside astrophysical BLs have 
been primarily concerned with the sub-sonic regime when the flow 
can be considered as almost incompressible \citep{KippenhahnThomas,
  Fujimoto1}. This
regime may indeed apply in well-developed BLs with smooth shear 
profiles, even though at the present level of knowledge one can 
hardly exclude the possibility of the existence of localized regions 
in steady-state BLs where compressibility is still important. 
Our primary goal here is to extend these studies into the
regime of highly compressible, supersonic flows and to explore the 
implications for astrophysical objects. In the course
of studying supersonic shear instabilities, we will sometimes 
also incorporate stratification in our calculations and explore 
the role of rotation.

The paper is organized as follows. In \S \ref{goveqn} we present our
governing set of equations and describe
the formalism we use to study the stability problem. 
Then in \S \ref{compress_KH}
we study the case of the compressible vortex sheet with no gravity
and show that our formalism reproduces the dispersion relation
obtained by previous authors \citep{Landau, Hatanaka, Pai, MilesKH, 
Gerwin}. In \S
\ref{compress_grav}, we introduce gravity as a small parameter and
perform a perturbation expansion to study what influence this has on
the stability of the vortex sheet. Finally, in \S \ref{finite_width_SL} we
study the stability of a finite width shear layer with a linear velocity
profile. We derive new dispersion relations for this case and test
the growth rate of the fastest growing mode using the
Godunov code Athena \citep{Stoneetal}.  

%%%%%%%%%%%%%%%%%%%%%%%%%%%%%%%%%%%%%%%%%%%%%%%

\section{Formalism}
\label{goveqn}

Here we discuss the formalism that we use for analytic calculations throughout
the paper. We adopt cylindrical
coordinates $(\cp, \theta, z)$ and for simplicity assume that all
equilibrium quantities are independent of $z$. Thus, for our purposes
gravity is given by $-g(\cp) \cphat$ ($g(\cp) > 0$), and we ignore the vertical
stratification of the disk. Such a setup does not allow for the
baroclinic instability, which would require $d \Omega/dz \ne 0$, but it does
contain the necessary ingredients for studying shear
instabilities. Given our assumptions, the equation of hydrostatic
equilibrium reads
\ba
\gb(\cp)\equiv g(\cp) - \cp \Omega^2(\cp)= -\frac{1}{\rho} \frac{d P}{d \cp},
\ea
where $\gb(\cp)$ is the effective gravity. The equilibrium density, 
pressure, and sound speed are given by $\rho(\cp)$, $P(\cp)$, 
and $s(\cp)$ respectively.

 Denoting the $(\cp, \theta, z)$ velocities by
$(u, v, w)$, using $v = \Omega \cp$, and assuming
 adiabaticity, the Euler equations in cylindrical coordinates are
\ba
\label{Eulerstart}
-\frac{\partial \rho}{\partial t} &=&\frac{1}{\cp}
\frac{\partial}{\partial \cp}(\rho \cp u) +
\frac{1}{\cp}\frac{\partial}{\partial \theta}(\rho v) +
\frac{\partial}{\partial z} (\rho w) \\
\frac{D u}{D t} &=& -\frac{1}{\rho}\frac{\partial
  P}{\partial \cp} + \frac{v^2}{\cp} - g \\
\frac{D v}{D t} &=& -\frac{1}{\rho \cp} \frac{\partial
  P}{\partial \theta} - \frac{u v}{\cp} \\
\frac{D w}{D t} &=& -\frac{1}{\rho}\frac{\partial P}{\partial z} \\
\label{Eulerend}
\frac{D \left(P\rho^{-\gamma}\right)}{D t}  &=& 0 \\
\frac{D}{D t} &\equiv& \frac{\partial}{\partial t} + u
\frac{\partial}{\partial \cp} + \frac{v}{\cp} \frac{\partial}{\partial
  \theta} + w \frac{\partial}{\partial z}
\ea
On top of the zeroth order equilibrium state, we consider infinitesimal
two-dimensional perturbations of the form $\delta f(\cp)
\exp[i(m \theta + k_z z - \omega t)]$,
where the $\delta$ denotes an
Eulerian perturbation. Starting from Equations
(\ref{Eulerstart})-(\ref{Eulerend}), the linearized first 
order equations are given by
\ba
\label{rotceq}
i \bo \delta \rho &=& \frac{1}{\cp} (\cp \rho \delta u)' + \frac{i m \rho}{\cp}
\delta v + i k_z \rho \delta w \\
i \bo \delta u &=& \frac{1}{\rho}\delta P' + \frac{\gb}{\rho} \delta \rho
- 2 \Omega \delta v \\
i \bo \delta v &=& \frac{i m}{\cp \rho} \delta
  P + 2B \delta u\\
i \bo \delta w &=& \frac{i k_z}{\rho} \delta P \\
\label{roteeq}
i \bo \delta \rho &=& -C_L \rho \delta u + \frac{i \bo}{s^2} \delta P, 
\ea 
where $\bo = \omega - m \Omega$ is the phase speed in the locally
corotating frame, $B = (\Omega + d(\cp \Omega)/d\cp)/2$ is Oort's B
constant, $C_L = \gamma^{-1} d \ln P/ d \cp - d \ln
\rho/d \cp$ is the Ledoux discriminant, and the primes 
denote differentiation with respect to
$\cp$. Note that in Equations (\ref{rotceq})-(\ref{roteeq}), we have
made Cowling's approximation by ignoring the first order perturbation
to the gravitational potential \citep{Cowling}. This means that we are
taking the self-gravity of the BL to be negligible. 

We can reduce the above set of linear equations to a pair of equations in
$\delta P$ and $\delta u$:
\ba
\label{deltaPeq}
i \left(\frac{\bo^2}{s^2} - k_z^2 - \frac{m^2}{\cp^2}\right) \delta P &=
&\rho \left(C_L \bo + \frac{2Bm}{\cp} \right) \delta u
+ \frac{\bo}{\cp} (\cp \rho \delta u)' \\
\label{deltaueq}
i\rho (\bo^2 - \gb C_L - \kappa^2) \delta u &=& \bo
\delta P' + \frac{ \bo \gb}{s^2} \delta P - \frac{2 \Omega
  m}{\cp} \delta P.
\ea
Here we have introduced the epicyclic frequency $\kappa^2 = 4 B \Omega$.

One generally expects the BL width $\delta_{BL}$ to be small compared to
the stellar radius, $\delta_{BL}\ll R_*$. Under this assumption, we
show in Appendix \ref{Reduction} 
that Equations (\ref{deltaPeq})-(\ref{deltaueq}) can be simplified 
to the following set of equations:
\ba
\label{deltaPeq1}
i \left(\frac{\bo^2}{s^2} - k_z^2 - \frac{m^2}{R_*^2}\right) \delta P &=
&\rho \left(C_L \bo + \frac{m}{R_*} S \right) \delta u
+ \bo (\rho \delta u)' \\
\label{deltaueq1}
i\rho (\bo^2 - \gb C_L) \delta u &=& \bo
\delta P' + \frac{ \bo \gb}{s^2} \delta P.
\ea
The form of Equations (\ref{deltaPeq1}) and (\ref{deltaueq1}) is the
same as for a plane-parallel stratified shear flow \citep{Alexakis}. 
This suggests that
the effects of Coriolis force and curvature are unimportant in a
radially thin BL if shear instabilities provide the turbulence, and
leads us to redefine the cylindrical $(\cp, \theta, z)$
coordinate system into a Cartesian $(x, y, z)$ system. The perturbed
quantities now have the form $\delta f(x)
\exp[i(k_y y + k_z z - \omega t)]$, where $k_y \equiv m/R_*$. From this it
immediately follows that $\bo  = \omega - k_y V_y(x)$, where $V_y(x) =
R_*\Omega(x)$ is the velocity profile. From here on, we will
  ignore the rotation terms, and treat the flow in the BL as a
  plane parallel shear flow.

From Squire's theorem for plane parallel shear flows \citep{Fejer},
any three-dimensional perturbation is mathematically equivalent to a
  two-dimensional perturbation upon making the transformations $k_y
  \rightarrow k_y/\cos \theta$, $k_z \rightarrow 0$, and $V_y \rightarrow
  V_y \cos \theta$, where $\cos \theta =
k_y/\sqrt{k_y^2+k_z^2}$. Thus, it is sufficient to only consider
two-dimensional perturbations ($k_z = 0$) in the stability problem,
and for the rest of this paper we take perturbations in the form
\ba
\label{pertform}
\delta f(x) \exp[i(k_y y - \omega t)].
\ea

Assuming two-dimensional
perturbations, Equations (\ref{deltaPeq1}) and (\ref{deltaueq1}) become
\ba
\label{deltaueq2}
i \left(\frac{\bo^2}{s^2} - k_y^2\right) \delta P &=
&\rho \left(C_L \bo + k_y V_y' \right) \delta u
+ \bo (\rho \delta u)' \\
\label{deltaPeq2}
i\rho (\bo^2 - \gb C_L) \delta u &=& \bo
\delta P' + \frac{ \bo \gb}{s^2} \delta P,
\ea
where the primes now denote differentiation with respect to
$x$. Equations (\ref{deltaPeq2}) and (\ref{deltaueq2}) can be used to
obtain the generalized Rayleigh equation (Appendix
\ref{gRapp}), which using a notation similar to \citet{Alexakis} reads
\ba
\label{gen_Rayleigh}
\delta \phi'' + \left(k_x^2 - g\tilde{\rho}\frac{k_s + k_g}{\tilde{W}^2} -
  \frac{\tilde{W}''}{\tilde{W}} \right) \delta \phi &=& 0.
\ea
Here, 
\ba
\label{deltaphi_eq}
\delta \phi = -\delta u
\sqrt{\rho}/k_x,
\ea
is a modified stream function, and for simplicity we have dropped the bar over
$\gb$, so now $g$ denotes the effective gravity. We have
also defined the quantities:
\begin{flalign}
\tilde{W} &= k_y W \sqrt{\ti{\rho}}/ik_x \\
W &= V_y - c \text{ where }  c = \omega/k \text{ is the phase speed.}
\\
\label{kxwavevec}
k_x^2 &= k_y^2(W^2/s^2-1) \text{ is the square of the x-component of
  the wavevector} \\
& \ \ \ \ \ \ \ \ \ \ \ \ \ \ \ \ \ \ \ \ \ \ \ \ \text{in the absence of shear or
  stratification.} \\
\tilde{\rho} &= \rho f^2 \text{, where $\rho$ is the density.} \\
f &= \exp\left(\int_0^x k_g(\zeta) d\zeta \right). \\
k_g &= g/s^2 \text{ is a measure of the inverse of the local scale height.} \\
k_s &= \rho'/\rho \text{ is the inverse stratification length scale.}
\end{flalign}
Note that the Ledoux discriminant is given by $C_L = -(k_g + k_s)$.
 
The boundary conditions on $\delta \phi$ that must be satisfied at a
discontinuity in the density or the velocity are that the upper and 
lower fluids stay in contact and that the pressure perturbation is 
continuous across the interface. We show explicitly in Appendix 
\ref{BCapp} that these conditions can be formulated as
\begin{eqnarray}
\label{BC_contact}
\frac{\delta \phi_+}{\ti{W}_+} &=& \frac{\delta \phi_-}{\ti{W}_-} \\
\label{BC_derivative}
\tilde{W}_+\delta \phi_+' - \tilde{W}_+'\delta \phi_+ - \frac{g_+ \tilde{\rho}_+
  \delta \phi_+}{\tilde{W}_+} &=& \tilde{W}_-\delta \phi_-' - \tilde{W}_-'\delta \phi_- - \frac{g_- \tilde{\rho}_-
  \delta \phi_-}{\tilde{W}_-}.
\end{eqnarray}
The $+$/$-$ signs denote evaluation of a
  quantity in the upper/lower fluid at the location of the
  interface.

%%%%%%%%%%%%%%%%%%%%%%%%%%%%%%%%%%%%%%%%%%%%%%

\section{The Vortex Sheet without Gravity}
\label{compress_KH}

Equations (\ref{gen_Rayleigh})-(\ref{BC_derivative}) are fully general
and apply to an arbitrary velocity profile. In particular, we can 
assume that the velocity varies discontinuously at some radius 
$x=0$:
\ba
\label{vsheet}
V_y(x) = \left\{
     \begin{array}{lr}
       \bar{V}_y, \ x > 0 \\
      -\bar{V}_y, \ x < 0,
     \end{array}
   \right.
\ea
where $\bar{V}_y$ is a constant. This is known as a 
vortex sheet approximation. It represents the simplest possible 
description of the velocity variation between the two limiting
values by essentially ignoring the details of the transition.
Subsequently in \S \ref{finite_width_SL}, we explore a more 
realistic model of the velocity variation, in which the
transition occurs in a region of finite radial width. We
point out that the vortex sheet approximation is valid
for $k_y \delta_{BL} \ll 1$. We show in Appendix \ref{KH_conv} that in this limit
the dispersion relation for a finite width layer of constant shear
reduces to the vortex sheet dispersion relation.

We assume in this section that $g=0$, so there is no gravity,
and that $\rho$ and $s$ are constant above and below the interface, but can be
discontinuous across it. This case has already been considered by other
authors in the past including \citet{Landau}, \citet{Hatanaka},
\citet{Pai}, \citet{MilesKH}, and \citet{Gerwin}. However, the results for the case
without gravity will often be referenced later in the paper and serve as a
verification of the formalism we developed in \S \ref{goveqn}. 

Using the velocity profile (\ref{vsheet}), the generalized Rayleigh 
equation (\ref{gen_Rayleigh}) becomes
\ba
\label{RCS}
\delta \phi_\pm'' + k_{x,\pm}^2 \delta \phi_\pm &=& 0, \\
\ea
where just as in \S \ref{goveqn}, the $+$/$-$ signs denote the upper/lower fluids respectively.
Since $k_{x,\pm}$ is constant for $V_y(x)$ given by Equation (\ref{vsheet}), 
we have that
\ba
\label{pert_eq}
\delta \phi_\pm &\propto& e^{-i k_{x,\pm} x}. 
\ea
In general, $k_{x,\pm}$ is complex, and the sign is determined by applying
the appropriate boundary conditions. We discuss the boundary
conditions shortly, which will also make clear the reason for the negative
sign in the exponential of Equation (\ref{pert_eq}). Plugging 
the expressions for $\delta \phi_\pm$ into Equation (\ref{BC_derivative}) 
gives 
\ba
\ti{W}_+ k_{x,+} \delta \phi_+ = \ti{W}_- k_{x,-} \delta \phi_- ,
\ea
where we have used $\delta \phi_\pm' = i k_{x,\pm} \delta \phi_\pm$,
$\ti{\rho}_\pm = \rho_\pm$, and $\ti{W}_\pm' = 0$. Using Equation (\ref{BC_contact})
to substitute for $\delta \phi_-$ in terms of $\delta \phi_+$, and
substituting for $\ti{W}$ in terms of $W$ and $\rho$ we have
\ba
\label{preprecompressdis}
\frac{k_y}{k_{x,+}}\rho_+W_+^2 = \frac{k_y}{k_{x,-}}\rho_-W_-^2.
\ea 
Introducing the density ratio $\epsilon = \rho_+/\rho_-$, the Mach
number in the upper fluid $M =
\bar{V}_y/s_+$, and the phase speed normalized by the sound speed in
the upper fluid $\varphi = c/s_+$, we have
\ba
\label{precompressdis}
\epsilon (M-\varphi)^2 \sqrt{(M+\varphi)^2 \frac{s_+^2}{s_-^2} - 1} =
(M+\varphi)^2 \sqrt{(M-\varphi)^2 - 1}.
\ea
By the definition of the sound speed, $s^2 = \gamma P/ \rho$, the
condition of pressure balance everywhere throughout the flow
requires that $\gamma_+^{-1}\rho_+s_+^2 = \gamma_-^{-1}\rho_-s_-^2$, 
where
$\gamma_+$ and $\gamma_-$ are the adiabatic indices above and below
the interface.
Assuming $\gamma_+ = \gamma_-$, we have from pressure balance that
$(s_-/s_+)^2 = \rho_+/\rho_- = \epsilon$. Thus, the
dispersion relation becomes:
\ba
\label{compressdis}
\epsilon (M-\varphi)^2 \sqrt{(M+\varphi)^2 \epsilon^{-1} - 1} =
(M+\varphi)^2 \sqrt{(M-\varphi)^2-1}.
\ea
\citet{MilesKH} has studied Equation (\ref{compressdis}) and found that
the stability criterion, i.e. that $\varphi$ is purely real, is 
given by 
\ba
M > M_\text{crit} = \frac{1}{2}(1 + \epsilon^{1/3})^{3/2}.
\label{Mcritunstable}
\ea
This shows the surprising result that infinitesimal disturbances are
stabilized at high Mach number. 

However, \citet{Fejer} pointed out
that due to Squire's theorem, it is always possible to choose an angle
$\theta$ for the wavevector with respect to the flow velocity such
that the projected Mach number $M \cos\theta$ is smaller than
$M_\text{crit}$. Thus, even at high Mach number, the vortex sheet
without gravity is still unstable to three dimensional
disturbances, which are almost perpendicular to the direction of the
flow; these unstable oblique modes resemble classical KH
modes. However, in the astrophysical context, a thin disk has a
scale height
$s/\Omega \ll R_*$, and the wavelength of the oblique modes will be small relative to
the disk scale height only for very small wavelengths $\lambda \ll
s/\Omega$. If the BL itself has a thickness $\delta_{BL} \gtrsim
s/\Omega$, the vortex sheet approximation for the oblique modes is
invalid, since either the modes don't fit into a disk
scale height, or the condition $\lambda \gg \delta_{BL}$ is not
satisfied.
%%%%%%%%%%%%%%%%%%%%%%%%%%%%%%%%%%%%%%%%%%%

\subsection{Solutions for $M \gg 1$}
\label{ratiomach}

Since we are interested in the high Mach number limit for the
initiation of the BL, we now find analytical solutions to 
the dispersion relation (\ref{compressdis}) for $M \gg M_\text{crit}$,
where $M_\text{crit}$ was defined in Equation (\ref{Mcritunstable}). 
Squaring Equation (\ref{compressdis}), one obtains a sixth order polynomial
in $\varphi$:
\ba
\label{polynomial}
(\epsilon^2 - (M+\varphi)^2 \epsilon)(M-\varphi)^4 -
(1-(M-\varphi)^2)(M+\varphi)^4 = 0.
\ea
This polynomial has two easy to find analytic factors \citep{MilesKH}
\ba
\label{tworoots}
\vp = \left \{-M\left(\frac{1-\eps^{1/2}}{1+\eps^{1/2}}\right),
-M\left(\frac{1+\eps^{1/2}}{1-\eps^{1/2}}\right) \right \}.
\ea
We now take the limit $M \gg M_\text{crit}$, in which case to terms of order
$\mathcal{O}(M^{-4})$, the other four solutions to the polynomial in Equation
  (\ref{polynomial}) are
\begin{multline}
\label{fourroots}
\vp = \left \{M+1+\frac{\eps}{2(2M+1)^2}, \ M-1-\frac{\eps}{2(2M-1)^2}, \right. \\
 -M + \eps^{1/2} + \frac{\eps^{1/2}}{2(2M-\eps^{1/2})^2}, \
\left. -M -
\eps^{1/2} - \frac{\eps^{1/2}}{2(2M+\eps^{1/2})^2} \right \}.
\end{multline}

It is easy to see that the six roots of the polynomial
(\ref{polynomial}) correspond to sound waves.
Starting from Equation (\ref{kxwavevec}) and rearranging terms we obtain
\ba
W^2 k_y^2 = s^2(k_x^2 + k_y^2).
\ea
As long as $k_x$ is real, this is the dispersion relation for a sound
wave, since $W^2$ is the square of the phase velocity in the frame
comoving with the fluid. Plugging in the six roots from Equations
(\ref{tworoots}) and
(\ref{fourroots}) into Equation (\ref{kxwavevec}), it is
straightforward to verify that $k_{x,\pm}$ are indeed real for all of
them.
Each of the four roots in Equation (\ref{fourroots}) has a further
simple
interpretation. The first two correspond to sound waves that propagate
almost parallel to the interface in the $+y$ and $-y$ directions in the upper
fluid, whereas the second two
correspond to sound waves that propagate almost parallel to the interface
in the $+y$ and $-y$
directions in the lower fluid. The two roots in Equation
(\ref{tworoots}) are more difficult to
interpret, but the first of these corresponds to a standing wave when
the two fluids have equal density (i.e. $\eps = 1$). 

Since each of the six roots for $M \gg M_\text{crit}$
yields a real value for $k_{x,\pm}$, the solutions do not damp
away from the interface, and we need to
apply radiation boundary conditions at $x=\pm \infty$. The
proper procedure is to demand
that all waves are outgoing in each fluid in a frame which is subsonic
with respect to the fluid \citep{Milessound}, and it is convenient to
work in the comoving frame of each fluid. The dimensionless phase
velocity in the frame comoving with the upper fluid is
$\varphi_{+,CF} = \varphi - M$, and the analogous expression for the
lower fluid is $\varphi_{-,CF} = \varphi + M$. 

In order to satisfy Equation (\ref{preprecompressdis}), $k_{x,+}$ and
$k_{x,-}$ must have the same sign, so we
must have either $\delta \phi_\pm \propto e^{i(|k_{x,\pm}| x - \omega
  t)}$ or $\delta \phi_\pm \propto e^{i(-|k_{x,\pm}| x - \omega
  t)}$. Moreover, since $k_{x,+}$ and $k_{x,-}$ have the same sign, it
is clear that $\varphi_{+,CF}$ and  $\varphi_{-,CF}$ must have the
opposite sign to yield outgoing waves in the comoving frames of each of
the two fluids.
Only three of the six roots found above satisfy this condition:
\ba
\label{lbranch}
\varphi_l &=& -M + \eps^{1/2} + \frac{\eps^{1/2}}{2(2M-\eps^{1/2})^2} \\
\label{mbranch}
\varphi_m &=& -M\left(\frac{1-\eps^{1/2}}{1+\eps^{1/2}}\right) \\
\label{ubranch} 
\varphi_u &=& M-1-\frac{\eps}{2(2M-1)^2}. 
\ea
We will refer to these three roots as the lower, middle,
and upper branches, respectively. Furthermore, it is straightforward
to check that the
solutions which yield outgoing waves in both the upper and lower
fluids have $\delta \phi_\pm \propto
e^{i(-|k_{x,\pm}| x - \omega t)}$, so $k_{x,\pm}$ are positive given
our definition (\ref{pert_eq}).

%%%%%%%%%%%%%%%%%%%%%%%%%%%%%%

\subsection{Dispersion Relation in the General Case}

\begin{figure}[!h]
  \centering
  \subfigure[]{\includegraphics[width=.49\textwidth]{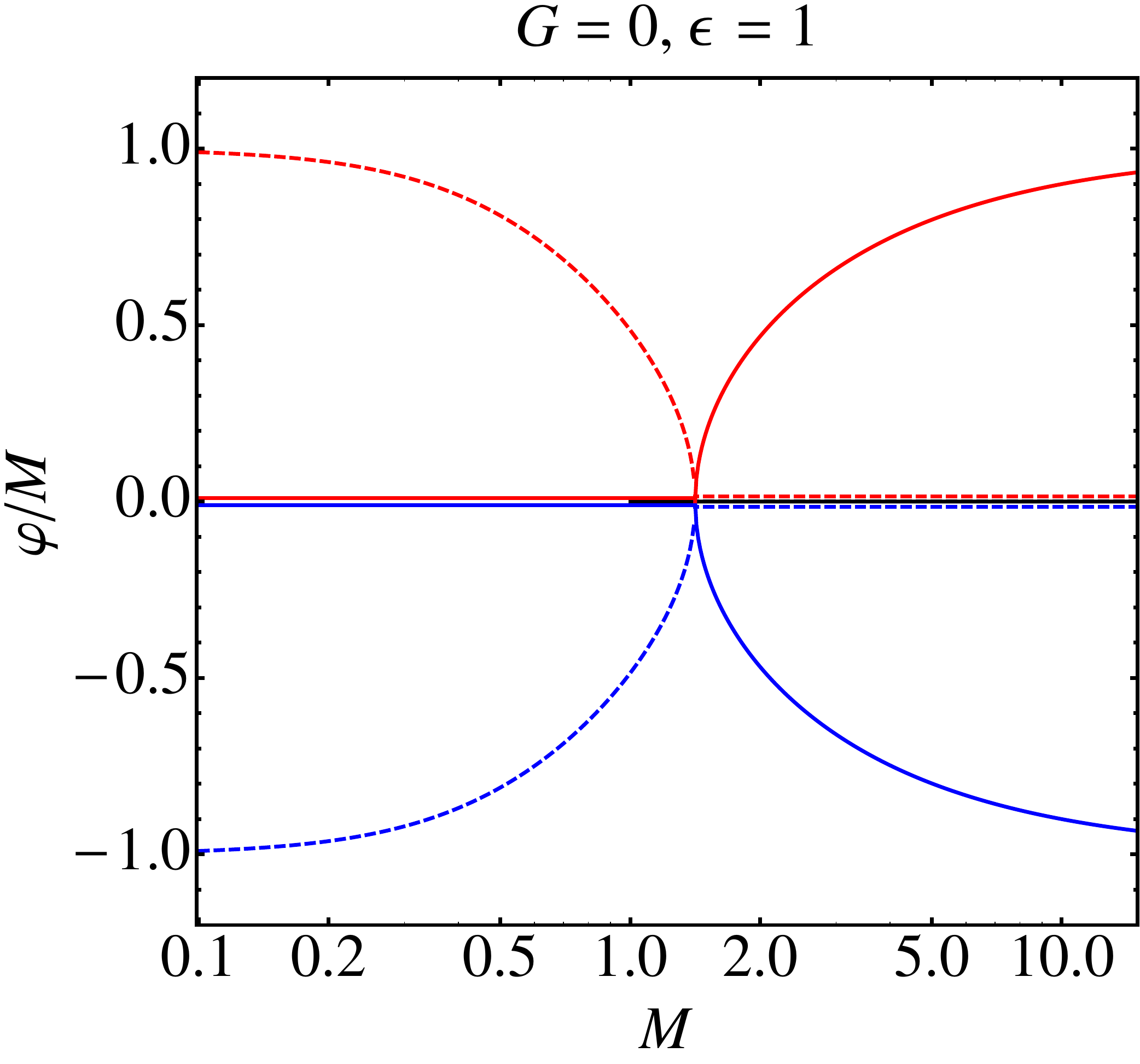}}
  \subfigure[]{\includegraphics[width=.49\textwidth]{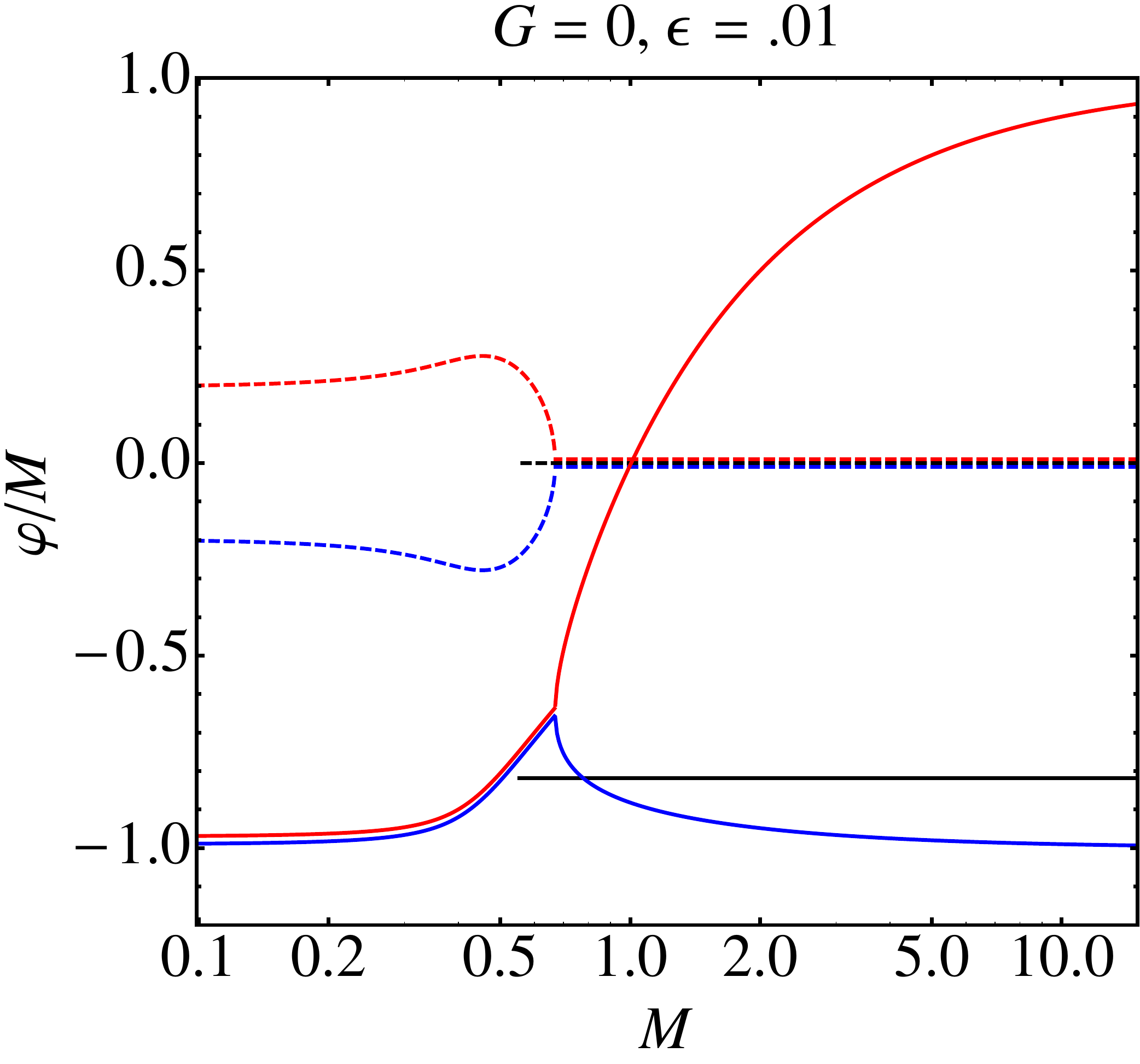}} 
  \caption{The dispersion relation $\varphi/M$ as a
  function $M$ for $\epsilon = 1$ and $\epsilon = .01$. The solid and
  dashed curves give the real and imaginary components respectively.
  The blue, black, and red curves
  correspond to the lower, middle, and upper branches
  respectively. The red and blue curves have been slightly offset
  vertically so they do not overlap.}
 \label{drfig}
\end{figure}

We now relax the assumption of $M \gg 1$ and numerically solve the dispersion
relation (\ref{compressdis}) at arbitrary Mach number for
$\epsilon = 1$ (Figure \ref{drfig}a) and $\epsilon = .01$ (Figure \ref{drfig}b). At the
critical Mach number given by Equation (\ref{Mcritunstable}), the
upper and lower branches merge together in the real plane and bifurcate in the
complex plate. These bifurcated solutions turn into the two
incompressible KH modes for $M \ll M_\text{crit}$.
Unlike, the lower and upper branches, the middle branch has no
incompressible analog and
ceases to be a viable physical solution below a critical Mach number
\ba
2M_m = 1 + \epsilon^\frac{1}{2}.
\label{Mcritmiddle}
\ea 
The reason for this is that below $M=M_m$, $k_{x,\pm}$ switches from real
to imaginary and the boundary conditions for the middle branch at $x =
\pm \infty$ can no longer be satisfied.

%%%%%%%%%%%%%%%%%%%%%%%%%%%%%%%%%%%%%%%%%%%%%%%%%%%%%

\section{The Isothermal Vortex Sheet with Gravity}
\label{compress_grav}

We now go beyond the simplifying assumption of no gravity used
in  \S \ref{compress_KH} and consider the case where $g$ is non-zero 
and constant. We shall shortly assume an isothermal equation of state,
but for now we consider the more general
polytropic equation of state of the form $P = K_\pm \rho^n$ in each of
the two fluids. For any equation
of state of this form, we have
\ba
k_g = \frac{g}{s^2} = -\frac{\frac{\partial P}{\partial \rho}\frac{d
    \rho}{dr}}{\gamma P} = - \frac{n}{\gamma}\frac{d \ln \rho}{dr} =
-\frac{n}{\gamma} k_s,
\ea
which means $k_s + k_g = (1-\gamma/n)k_g$. If $n = \gamma$,
corresponding to the case of an adiabatic atmosphere, the
second term in parentheses in Equation (\ref{gen_Rayleigh}) drops out,
but then the third term becomes difficult to treat
analytically. However, for an isothermal atmosphere ($n=1$) the sound speed
is constant, and both the second and third terms in Equation
(\ref{gen_Rayleigh}) reduce to a tractable form. Since we are only interested 
in the gross, qualitative properties of the flow, we 
assume both fluids are isothermally stratified, since this
assumption significantly simplifies the analytical treatment.

For $n = 1$, the
second term in Equation (\ref{gen_Rayleigh}) becomes
\ba
-g\tilde{\rho}\frac{k_s + k_g}{\tilde{W}^2} &=& g k_g (1-\gamma)
\frac{k_x^2}{k_y^2 W^2}
\label{isosecondterm} 
\ea
We now again assume a vortex sheet velocity profile (Equation
(\ref{vsheet})), in which case for $n=1$, the third term in Equation 
(\ref{gen_Rayleigh})
is given by
\ba
\label{isothirdterm}
\frac{\tilde{W}''}{\tilde{W}} =
\frac{(\sqrt{\tilde{\rho}})''}{\sqrt{\tilde \rho}} = \left(\frac{2-\gamma}{2}k_g\right)^2
\ea
Putting Equations (\ref{isosecondterm}) and (\ref{isothirdterm}) into
Equation (\ref{gen_Rayleigh}), we have
\ba
&&\delta \phi'' + \ti{k}_x^2 \delta \phi = 0, \nonumber\\
&& \ti{k}_x^2\equiv\left[k_x^2
  \left(1-(\gamma-1) \left(\frac{k_g}{k_y}\right)^2
  \left(\frac{s}{W}\right)^2\right) - \left(\frac{2-\gamma}{2}
  k_g\right)^2 \right]. 
\label{Rayleigh_isothermal}
\ea
The perturbation is then given as 
\ba
\label{pertwithgrav}
\delta \phi_\pm \propto e^{-i
\ti{k}_{x,\pm} x}.
\ea

We now comment on the terms present in Equation
(\ref{Rayleigh_isothermal}). In the absence of gravity, $k_g = 0$,
and we simply have $\ti{k}_x = k_x$. If $k_g \ne 0$, then the second
term on the right hand side of Equation (\ref{Rayleigh_isothermal})
can be written in a more familiar form as
\ba
(\gamma-1) \left(\frac{k_g}{k_y}\right)^2 \left(\frac{s}{W}\right)^2 =
\left(\frac{N}{\omega - k_yV_y}\right)^2,
\ea
where 
\ba
\label{eqBrunt}
N^2 = \frac{(\gamma - 1) g^2}{s^2}
\ea
is the Brunt-V\"{a}is\"{a}l\"{a} frequency. Thus, the second term on
the right hand side of Equation (\ref{Rayleigh_isothermal}) provides a
lower frequency cutoff for sound waves at
the Brunt-V\"{a}is\"{a}l\"{a} frequency. The last term in Equation
(\ref{Rayleigh_isothermal}) arises because we have {\it not} made the short
wavelength approximation $k_x \gg k_g$.

Now, we use the boundary conditions (\ref{BC_contact}) and
(\ref{BC_derivative}) to determine the dispersion relation. 
Substituting $\delta \phi_\pm' = -i \ti{k}_{x,\pm} \delta \phi_\pm$, and
$\ti{W}_\pm' = (2-\gamma) k_{g,\pm} \ti{W}_\pm/2$ into the condition 
(\ref{BC_derivative}), we have 
\ba
-\left[i\ti{k}_{x,+} + \left(\frac{2-\gamma}{2}
\frac{g}{s_+^2}\right)\right] \ti{W}_+ \delta \phi_+ - \frac{g \ti{\rho}_+ \delta \phi_+}{\ti{W}_+}
= -\left[i\ti{k}_{x,-} + \left(\frac{2-\gamma}{2}
\frac{g}{s_-^2}\right)\right]\ti{W}_-\delta \phi_- - \frac{g \ti{\rho}_- \delta \phi_-}{\ti{W}_-}.
\ea
Next, using the second boundary condition (\ref{BC_contact}) to substitute for
$\delta \phi_-$ in terms of $\delta \phi_+$, and using the fact that $\ti{\rho}_\pm
= \rho_\pm$ at the interface, we have
\ba
-\left[i\ti{k}_{x,+} + \left(\frac{2-\gamma}{2}
\frac{g}{s_+^2}\right)\right] \ti{W}_+^2 - g \rho_+
= -\left[i\ti{k}_{x,-} + \left(\frac{2-\gamma}{2}
\frac{g}{s_-^2}\right)\right]\ti{W}_-^2 - g \rho_-.
\ea
We now introduce the dimensionless gravity parameter
\ba
G = \frac{g}{k_y s_+^2},
\label{eq:grav}
\ea
which is closely related to the ratio
of the wavelength to the pressure scale height, $h_s$. Thus, $G \sim 1$ when
$k_y h_{s,+}
\sim 1$, and $G \eps^{-1} \sim 1$ when $k_y h_{s,-} \sim 1$.
Using $\epsilon =
\rho_+/\rho_- = s_-^2/s_+^2$, and performing some algebra, we have
\ba
\left[i\frac{\ti{k}_{x,+}}{k_y} + \left(\frac{2-\gamma}{2}G\right)\right]
\left(\frac{k_y}{k_{x,+}}\right)^2\left(\frac{{W}_+}{s_+}\right)^2\epsilon + G
(1-\epsilon)
= \left[i\frac{\ti{k}_{x,-}}{k_y} + \left(\frac{2-\gamma}{2}G
\epsilon^{-1}\right)\right]\left(\frac{k_y}{k_{x,-}}\right)^2\left(\frac{{W}_-}{s_+}\right)^2.
\label{mastergravity}
\ea
Setting $G=0$ it is clear that we recover the vortex sheet dispersion
relation in the absence of gravity (Equation (\ref{preprecompressdis})).

We now check Equation (\ref{mastergravity}) by showing
that it reproduces the well-known incompressible KH dispersion
relations in the limit
$M \ll M_\text{crit}$, before going on to treat the case of the supersonic vortex sheet with
gravity.

%%%%%%%%%%%%%%%%%%%%%%%%%%%%%%%%%%%%%%%%%%%%%%%%%%%%%

\subsection{Highly Subsonic Vortex Sheet with Gravity}
\label{subsonicKH}

We assume that $\bar{V}/s_\pm \ll 1$, which immediately implies $M \ll
M_\text{crit}$, and we also assume $G \ll 1$ and $G\epsilon^{-1} \ll 1$, which
means that the wavelength of the perturbation is much smaller than the
scale height in both the upper and lower fluids ($k_{g,\pm}/k_y \ll
1$). Next, we eliminate
sound wave modes by assuming
$\varphi \ll 1$ (phase velocity much lower than sound
velocity in upper fluid) and  $\varphi \epsilon^{-1/2} \ll 1$ (phase
velocity much lower than sound velocity in lower fluid). According to
Equation (\ref{Rayleigh_isothermal}), this means $k_x = \pm i k_y$,
and given our definition for $\delta \phi_\pm$ in Equation
(\ref{pertwithgrav}), we must chose the minus sign in the upper fluid
and the plus sign in the lower fluid to give vanishing solutions at $x =
 \pm \infty$.
Equation (\ref{mastergravity}) then yields
\ba
\label{subsonicdispers}
-\epsilon \left(\frac{\omega}{k_y}-\bar{V_y}\right)^2\sqrt{1-\frac{(\gamma-1)G^2}{(M-\varphi)^2}} +
\frac{g}{k_y}(1-\epsilon) = \left(\frac{\omega}{k_y} + \bar{V_y}\right)^2\sqrt{1-\frac{(\gamma-1)G^2\epsilon^{-1}}{(M+\varphi)^2}}.
\ea
Next, we note that $(\gamma-1)G^2/(M-\varphi)^2 =
N_+^2/(\omega-k_y\bar{V_y})^2$
and that $(\gamma-1)G^2\epsilon^{-1}/(M+\varphi)^2 =
N_-^2/(\omega+k_y\bar{V_y})^2$, where $N$ is the
Brunt-V\"{a}is\"{a}l\"{a} frequency for an isothermal atmosphere and
was given in Equation (\ref{eqBrunt}). We
expect $|\omega \pm k_y \bar{V}_y| \gtrsim \sqrt{gk}$, which is the
characteristic frequency of surface gravity waves. Since $N \pm \sim
\sqrt{g k_{g,\pm}}$, and we have already assumed $k_{g,\pm}/k_y \ll
1$, it follows that $N_\pm \ll
|\omega \mp k_y\bar{V_y}|$. Consequently, Equation
(\ref{subsonicdispers}) reduces to
\ba
\label{KHwellknown}
-\epsilon \left(\frac{\omega}{k_y} - \bar{V_y}\right)^2 +
\frac{g}{k_y}(1-\epsilon) = \left(\frac{\omega}{k_y} + \bar{V_y}\right)^2,
\ea
which is the well known KH dispersion relation for an incompressible
fluid in the presence of gravity.

%%%%%%%%%%%%%%%%%%%%%%%%%%%%%%%%%%%%%%%%%%%%%%%%%%%%%

\subsection{The Weak Gravity Limit at High Mach Number}
\label{weakgravity}

As mentioned before in \S \ref{ratiomach}, \citet{MilesKH} has 
demonstrated that in the absence of gravity,  
the vortex sheet is stable above a critical Mach number given by 
Equation (\ref{Mcritunstable}). We now address the question of whether 
the system still remains stable when $G \ne 0$.

To answer this question, we will use our general dispersion relation 
(\ref{mastergravity}) in which we will additionally assume $M \gg
M_\text{crit}$, since
this assumption significantly simplifies the algebra. Since we have 
already obtained solutions for the case $G=0$ and $M \gg M_\text{crit}$ 
(\S \ref{ratiomach}), we proceed by considering gravity as a
perturbation. We consider wavelengths that are much smaller than the
pressure scale height and ask what happens in the limit
$G \rightarrow 0$. In our analysis, we will assume that the density
ratio $\epsilon \le 1$, so that the system is
stable to the Rayleigh-Taylor instability.

Because the introduction of gravity makes the upper and lower fluids
stratified, some care should be taken when determining which solutions
are physical and which are not. \citet{Milessound} has shown that for
$G=0$, the three supersonic KH modes can be understood in terms of sound
waves emitted from the interface between the two fluids. This
interpretation is useful as well for the case with gravity and
leads to a couple of insights. First, the amplitude of sound waves
is not constant as they propagate through a stratified medium. Rather,
to conserve energy, waves propagating upward (to lower densities) must
increase in amplitude, and those propagating downward (to higher densities) must
decrease in amplitude. Second, since the sound waves are emitted from the interface,
if $\omega$ has an imaginary component, then the amplitude of the
emitted waves changes with time. Thus, if $\Im[\omega] > 0$, and
there is an instability, then the amplitude of the waves will decay
with distance from the interface, since the waves emitted in the past had lower
amplitude. Conversely if $\Im[\omega] < 0$ and the perturbation is
damping in time, then the amplitude of the emitted waves will increase with
distance from the interface, since the waves emitted in the past had a
larger amplitude. Both of these effects mean that the amplitude of the
waves can blow up as we move away from the interface. Thus, we take as
physical those solutions
which yield outgoing waves (away from the interface) in both the upper
and lower fluids, even if
these solutions diverge as $x \rightarrow \pm \infty$. For the limit
$G \rightarrow 0$, this means that the physical solutions are
still the lower, middle, and upper branches but now modified by the presence
of a weak gravitational field. 

We note here that although the dispersion relation
(\ref{mastergravity}) is valid for all values of $G$ and not just for
small $G$, the simple picture of outgoing sound waves in the upper and
lower fluids is only valid for $G \rightarrow 0$. From a physical
point of view, this can be attributed to the following fact. Sound waves
(p-modes) traveling in a stratified atmosphere have a frequency
cutoff at $\omega^2 = N^2$ below which propagation is not
possible. Given our definitions of $k_y$ and
$\ti{k}_x$, the frequency of a sound wave in the frame comoving with
the fluid is $\omega^2 \sim (k_y^2 + \ti{k}_x^2) s^2$. Substituting
$N^2 \sim \omega^2$ and using the definition of $N^2$
(Equation (\ref{eqBrunt})) yields 
$G^2<(\gamma-1)^{-1}(1 + (|\ti{k}_x|/k_y)^2)$ for sound waves to propagate. 
If this condition is not fulfilled, then the
picture of outgoing sound waves is invalid, and the boundary
conditions need to be formulated in a different way, which is beyond
the scope of the present work. 

%%%%%%%%%%%%%%%%%%%%%%%%%%%%%%%%%%%%%%%%%%%%%%%%%%%%%

\subsubsection{The Lower Branch}
\label{lbsec}

We begin by considering how the lower wave is modified in the limit $G \rightarrow
0$ and $M \gg M_\text{crit}$. If $G = 0$ exactly, then $\ti{k}_x = k_x$, and in the limit $G
\rightarrow 0$ we have $\ti{k}_x = k_x
\left(1 + \mathcal{O}(G^2)\right)$. Keeping terms only to first order
in $G$, Equation (\ref{mastergravity}) becomes
\ba
\left[i\frac{k_{x,+}}{k_y} + \left(\frac{2-\gamma}{2}G\right)\right]
\left(\frac{k_y}{k_{x,+}}\right)^2\left(\frac{{W}_+}{s_+}\right)^2\epsilon + G
(1-\epsilon)
= \left[i\frac{k_{x,-}}{k_y} + \left(\frac{2-\gamma}{2}G
\epsilon^{-1}\right)\right]\left(\frac{k_y}{k_{x,-}}\right)^2\left(\frac{{W}_-}{s_+}\right)^2.
\label{mastergravitysimple}
\ea
Writing this out explicitly in terms of $M$ and $\vp$ gives
\ba
\label{l0eq}
\frac{i(M-\varphi)^2 \eps}{\sqrt{(M-\vp)^2-1}} - \frac{i (M+\varphi)^2}{\sqrt{(M+\vp)^2\eps^{-1}-1}} +
G(1-\epsilon) =
\frac{2-\gamma}{2}G\left(\frac{(M+\varphi)^2
  \epsilon^{-1}}{(M+\varphi)^2\epsilon^{-1}-1} - \frac{(M-\varphi)^2\epsilon}{(M-\varphi)^2-1}
\right).
\ea

Next, we assume that gravity only weakly affects the
dispersion relation and make the perturbative expansion 
\ba
\label{phiseries}
\varphi = \varphi_0 + \varphi_1, \ \
|\varphi_1|/|\varphi_0| \ll 1,
\ea 
where $\vp_0$ is the solution for $G=0$ and $M \gg M_\text{crit}$.

For the lower branch, we can use Equation (\ref{lbranch}) for $\vp_0$, and in the
limit $M \gg M_\text{crit}$ we have
\ba
\varphi_0 &\approx& -M + \eps^{1/2} \\
\label{lkpeq}
\frac{k_{x,+}}{k_y} &=& \sqrt{(M-\varphi_0)^2-1} \approx
2M-\eps^{1/2} \\
\frac{k_{x,-}}{k_y} &=& \sqrt{(M+\varphi_0)^2\epsilon^{-1} -1} \approx \frac{1}{2M-\eps^{1/2}}.
\ea 

Defining 
\ba
\label{lmstar}
M_l \equiv 2M - \eps^{1/2},
\ea
Equation (\ref{l0eq}) becomes 
\ba
\label{l1eq}
\frac{i(M_l-\varphi_1)^2 \eps}{\sqrt{(M_l-\vp_1)^2-1}} - \frac{i
  (\eps^{1/2}+\varphi_1)^2}{\sqrt{(\eps^{1/2}+\vp_1)^2\eps^{-1}-1}} +
G(1-\epsilon) =
\frac{2-\gamma}{2}G\left(\frac{(\eps^{1/2}+\varphi_1)^2
  \epsilon^{-1}}{(\eps^{1/2}+\varphi_1)^2\epsilon^{-1}-1} -
  \frac{(M_l-\varphi_1)^2\epsilon}{(M_l-\varphi_1)^2-1}
\right).
\ea
It will turn out (Equation (\ref{lvp1eq})) that $\vp_1$ is proportional to
  $G$, so working to first order in $G$ is equivalent to working to
  first order in $\vp_1$. Equation (\ref{l1eq}) then simplifies to
\ba
i (M_l-\varphi_1) \epsilon - \frac{i M_l \eps (1 +
  2\eps^{-1/2} \vp_1)}{1+ M_l^2\eps^{-1/2} \vp_1} +
G(1-\epsilon) = \frac{2-\gamma}{2}G\left(M_l^2\frac{1+2
  \eps^{-1/2} \vp_1
  }{1 + 2 M_l^2 \epsilon^{-1/2}\vp_1} - \eps \right).
\ea
Assuming $|M_l^2 \epsilon^{-1/2}\vp_1| \ll 1$,
and keeping terms to leading order in $M_l$, we can solve for $\vp_1$
  in terms of $G$.
\ba
\label{lvp1eq}
\vp_{1,l}
\approx - \frac{2-\gamma}{2}
\frac{G}{M_l\eps^{1/2}} i.
\ea
We immediately
see two things from Equation (\ref{lvp1eq}). First, $\vp_1$ is
purely imaginary, and second, if $\gamma < 2$, $\vp_1$ is
negative and the perturbation damps, whereas if $\gamma > 2$,
$\vp_1$ is positive and the perturbation grows. For realistic
equations of state, $\gamma \le 5/3$, so the lower wave always damps.

%%%%%%%%%%%%%%%%%%%%%%%%%%%%%%%%%%%%%%%%%%%%%%%%%%%%%

\subsubsection{The Middle and Upper Branches}

We can find an approximate solution for $\vp_1$ in the limit $G
\rightarrow 0$ and $M \gg M_\text{crit}$ for the middle and upper branches in much the same
manner as for the lower branch. The first order correction for the
middle branch is
\ba
\label{vp1meq}
\vp_{1,m} = -\frac{\gamma}{2}\frac{G(1-\eps^{1/2})}{\eps^{1/2}}i,
\ea
and for the upper branch is
\ba
\label{uvp1}
\vp_{1,u}
\approx \frac{2-\gamma}{2}\frac{G \eps^{1/2}}{2M-1}i.
\ea

Just as in the case of the lower branch, $\vp_1$ is purely
imaginary for both the middle and upper branches. We see that
$\vp_1$ is negative for the middle branch if $\eps < 1$. However, for the
upper branch if $\gamma < 2$, $\vp_1$ is positive
and if $\gamma > 2$, $\vp_1$ is negative. This is opposite from the
lower branch meaning that for any $\gamma \neq 2$ in the limit
$M \gg 1$, one of the
two branches is always unstable and the other one is damped. For a
realistic equation of state, $\gamma \le 5/3$ {\it so the upper branch is
the unstable one}, and the lower
branch damps. 

\begin{comment}
One may wonder whether there are any completely new
modes introduced by the presence of gravity that have no analog in the
case $G=0$. As we show in Appendix \ref{completenessapp}, there are only three physically viable
solutions in the limit $G \rightarrow 0$, corresponding to the lower,
middle, and upper branches respectively.
\end{comment}

%%%%%%%%%%%%%%%%%%%%%%%%%%%%%%%%%%%%%%%%%%%%%%%%%%%%%

\subsubsection{Numerical Verification}

\begin{figure}[!h]
  \centering
  \subfigure[]{\includegraphics[width=.32\textwidth]{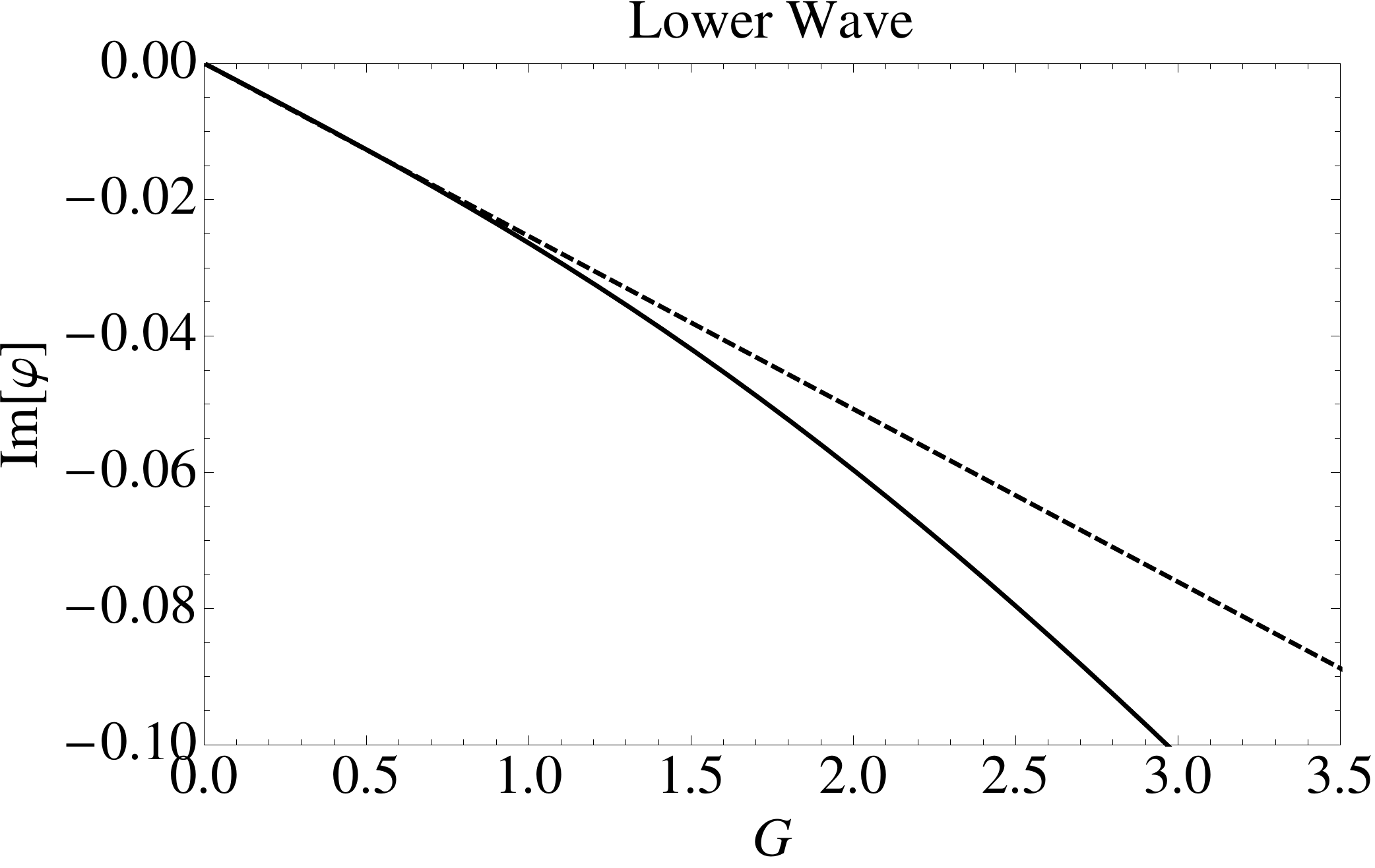}}
  \subfigure[]{\includegraphics[width=.32\textwidth]{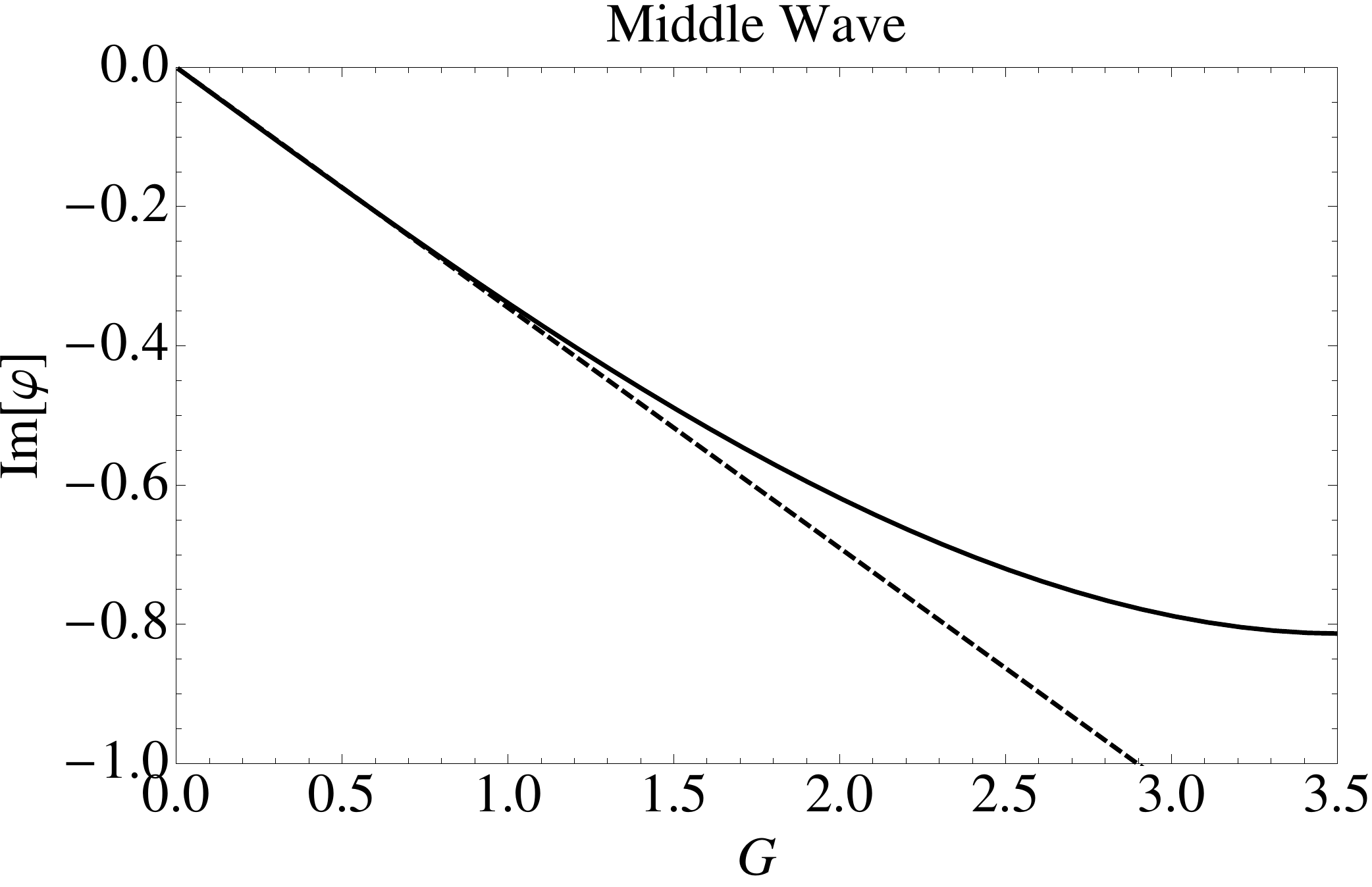}} 
  \subfigure[]{\includegraphics[width=.32\textwidth]{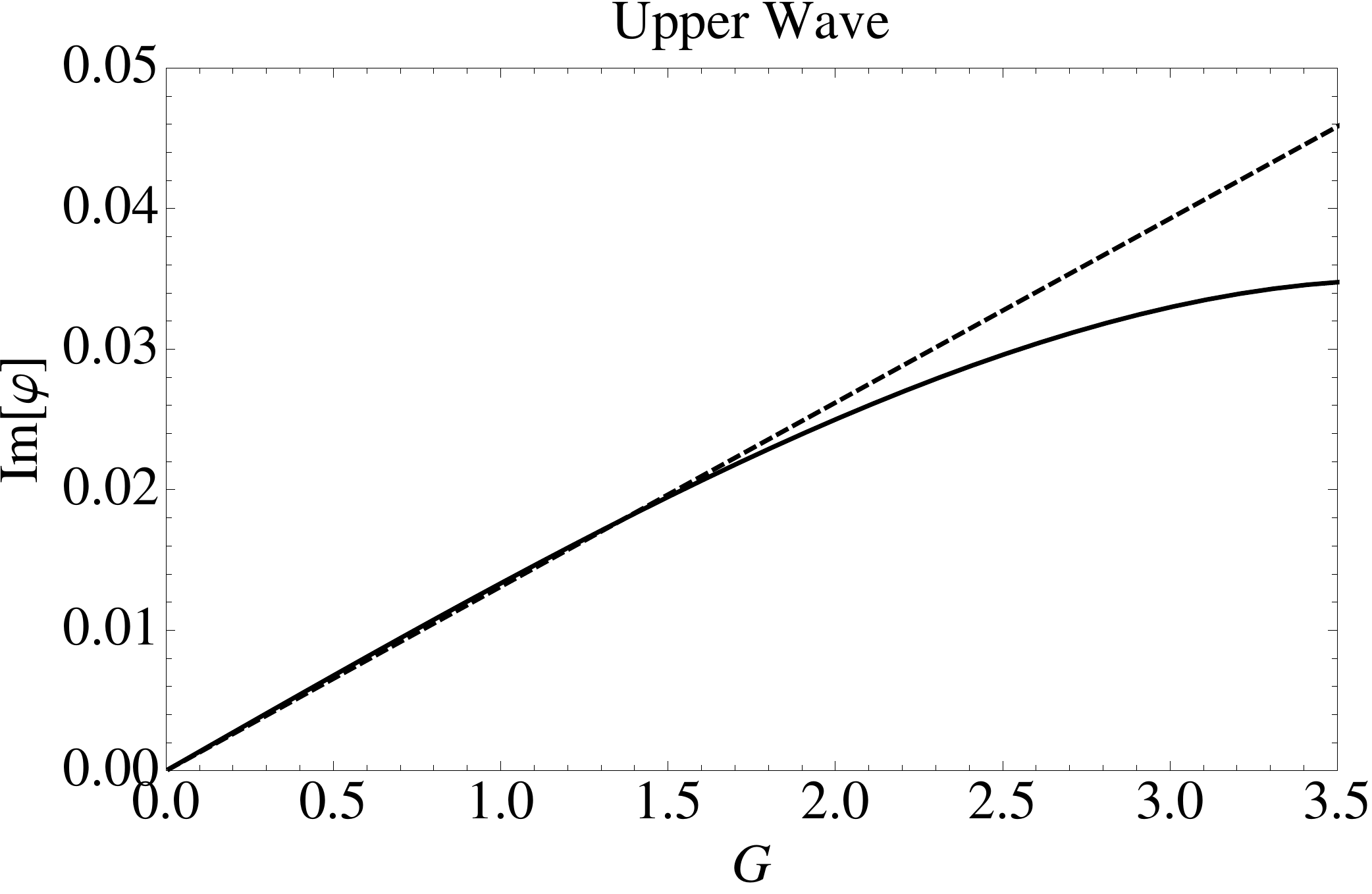}}
  \caption{Plots of $\Im(\varphi)$ vs. $G$ for the lower (a), middle
  (b), and upper (c) branches respectively, using $M=5$, $\eps=.5$,
  $\gamma = 5/3$. The solid curves show the
  numerical solution obtained by solving Equation
  (\ref{mastergravity}) and the dashed curves correspond to the
  approximate solutions from Equations (\ref{lvp1eq}), (\ref{vp1meq}),
  and (\ref{uvp1}).}
  \label{modefig}
\end{figure}

We verify Equations (\ref{lvp1eq}), (\ref{vp1meq}), and (\ref{uvp1})
numerically by solving the fully general
dispersion relation (\ref{mastergravity}) and comparing $\Im(\varphi)$
with our analytical estimate for the parameters $M=5$, $\eps = .5$,
and $\gamma = 5/3$. We plot $\Im(\varphi)$ vs. $G$ in
for both our analytical solutions (solid lines) and the ones obtained
numerically (dashed lines) in Figure \ref{modefig}. The analytical
solution converges to the numerical one in the limit $G \rightarrow 0$.

%%%%%%%%%%%%%%%%%%%%%%%%%%%%%%%%%%%%%%%%%%%%%%%%%%%%%

\section{Sonic Instabilities in a Finite Width Layer of Constant Shear}
\label{finite_width_SL}

The calculations presented in \S \ref{compress_KH},\ref{compress_grav} 
were performed for the velocity profile (\ref{vsheet}) featuring 
a discontinuity at some radius. We now consider a more complicated 
(and more realistic at later stages of the BL evolution)
initial setup in which the velocity between the two fluids varies 
{\it continuously} within a narrow shear layer. Unlike the vortex
sheet, the finite width shear layer without gravity is known
to be unstable at high Mach number
\citep{Glatzel,ChoudhuryLovelace,Ray}, and in this case the growth rate of the
instability scales inversely with the width of the shear layer and in
proportion to the shear, $S \sim \Omega_K R_*/\delta_{BL}$. 

In the following, we extend some of the findings of \citet{Glatzel} 
to study sonic instabilities in a finite width shear layer without 
gravity and apply them to the problem of the BL initiation. 
The setup we consider has the velocity profile
\ba
\label{vsetup}
V(x) = \left\{
     \begin{array}{lr}
       \bar{V}, \ x > \delta_{BL}, \\
       \bar{V}x/\delta_{BL}, \ -\delta_{BL} \le x \le \delta_{BL}, \\
       -\bar{V}, \ x < -\delta_{BL}
     \end{array}
   \right.
\ea
and the density profile
\ba
\label{rhosetup}
\rho(x) = \left\{
     \begin{array}{lr}
       \rho_+ , \ x > -\delta_{BL} \\
       \rho_-, \ x < -\delta_{BL}
     \end{array}
   \right.
\ea
Pressure equilibrium again requires that $\rho_+ s_+^2 = \rho_-
s_-^2$, which sets the sound speed everywhere in the flow, and as
before we have $\eps = \rho_+/\rho_-$ and $M = \bar{V}/s_+$. 

Although we only consider a linearly varying velocity
profile in the shear layer, \citet{Ray} and 
\citet{ChoudhuryLovelace} have found that different 
shear profiles are qualitatively similar. Thus, we 
consider a constant shear to be representative of more 
general shear profiles.

%%%%%%%%%%%%%%%%%%%%%%%%%%%%%%%%%%%%%%%%%%%%%%

\subsection{Dispersion Relation}

We now study the dispersion relation of the finite width shear
layer. In applying the dispersion relation to the initiation of the
BL, we are most interested in the
growth rate of the fastest growing mode for $M \gg 1$. 
\citet{Glatzel} has already obtained the dispersion
relation for the case $\eps = 1$, and using his techniques, we
derive the dispersion relation for the case of arbitrary $\epsilon$ in
Appendix \ref{Glatzelapp}. We also show in Appendix \ref{KH_conv}
that the dispersion relation for a finite width shear layer reduces
to the dispersion relation for a vortex sheet in the limit $k_y
\rightarrow 0$ at constant $\delta_{BL}$.

\begin{figure}[!p]
  \centering
  \subfigure{\includegraphics[width=.32\textwidth]{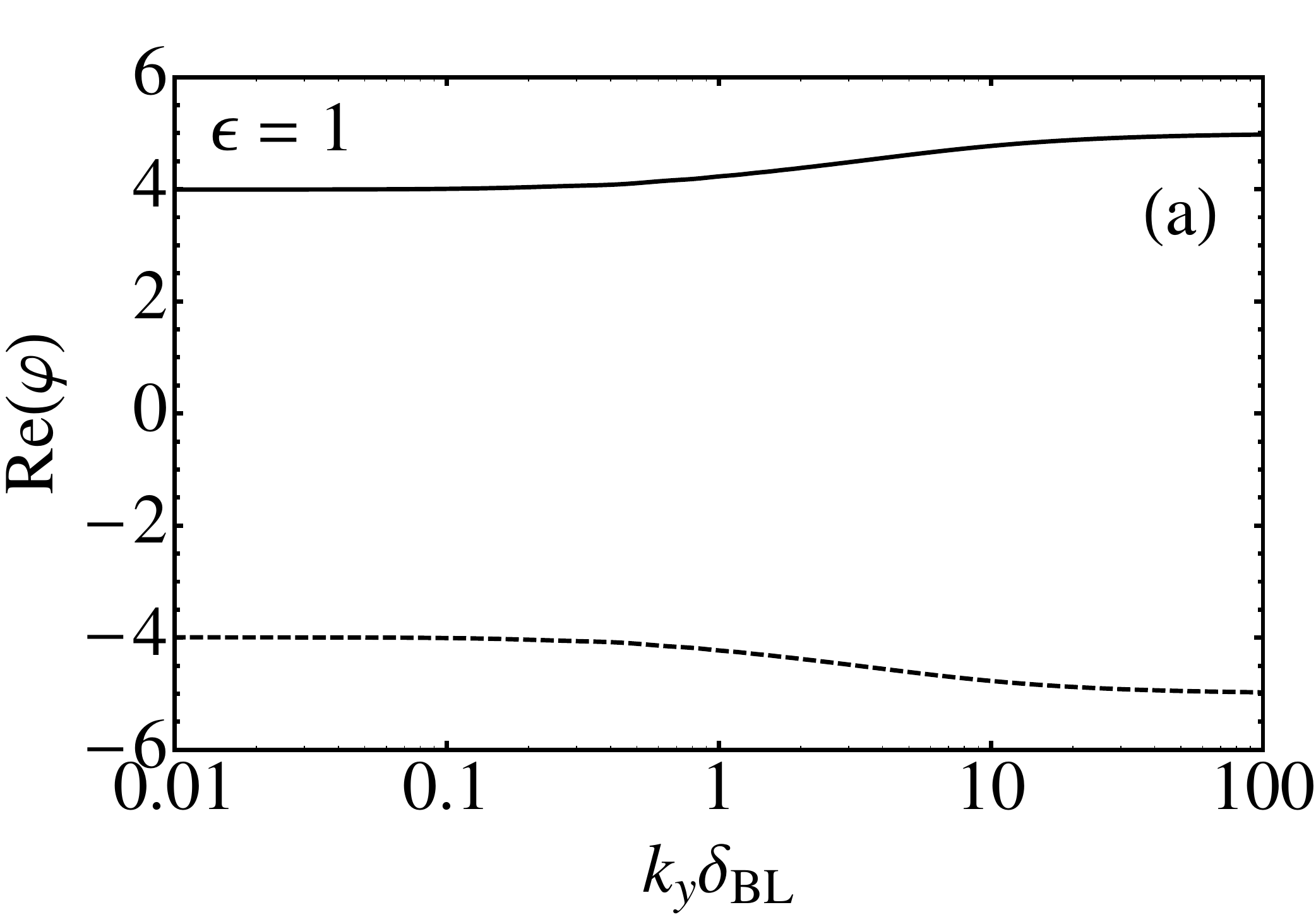}}
  \subfigure{\includegraphics[width=.32\textwidth]{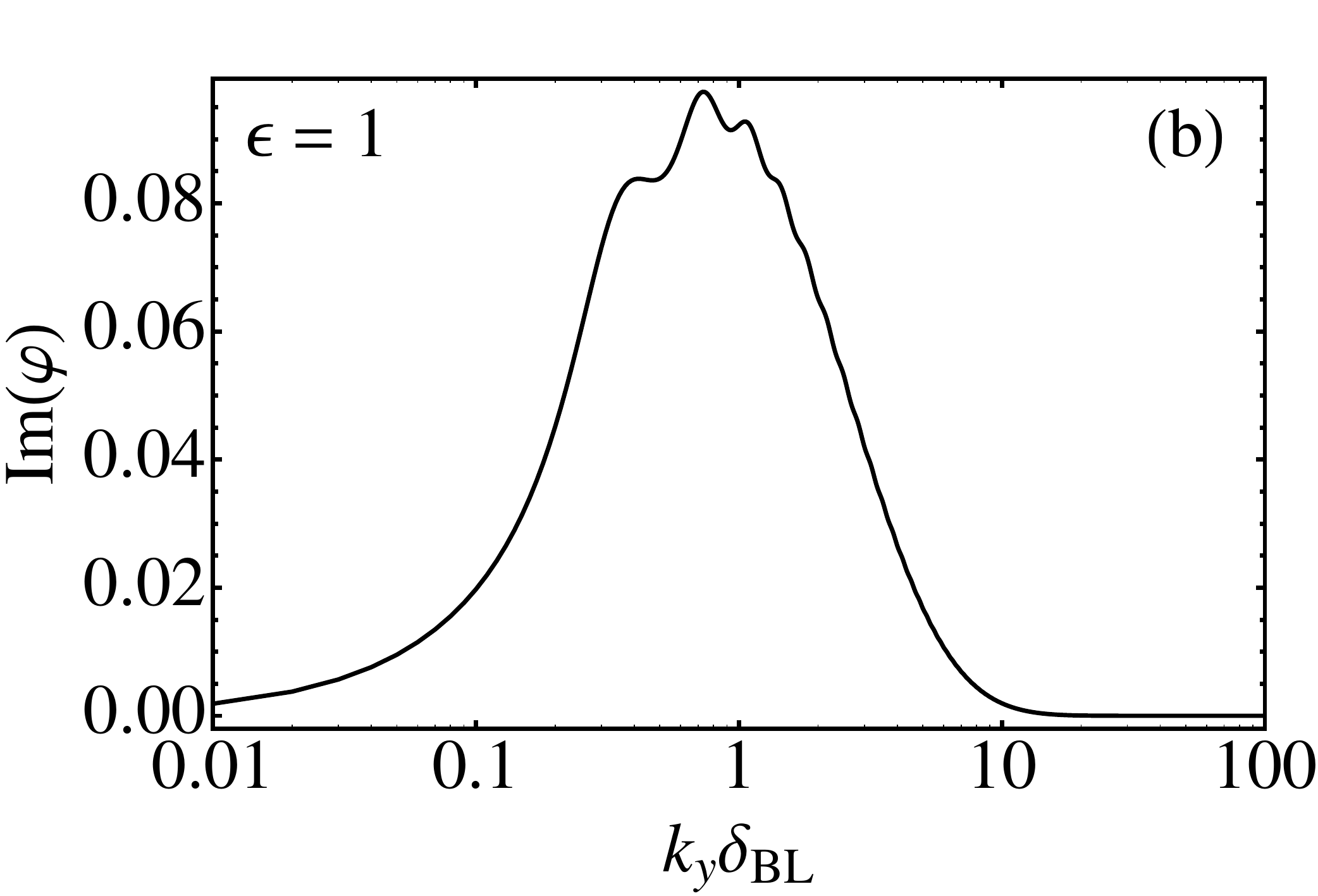}}
  \subfigure{\includegraphics[width=.32\textwidth]{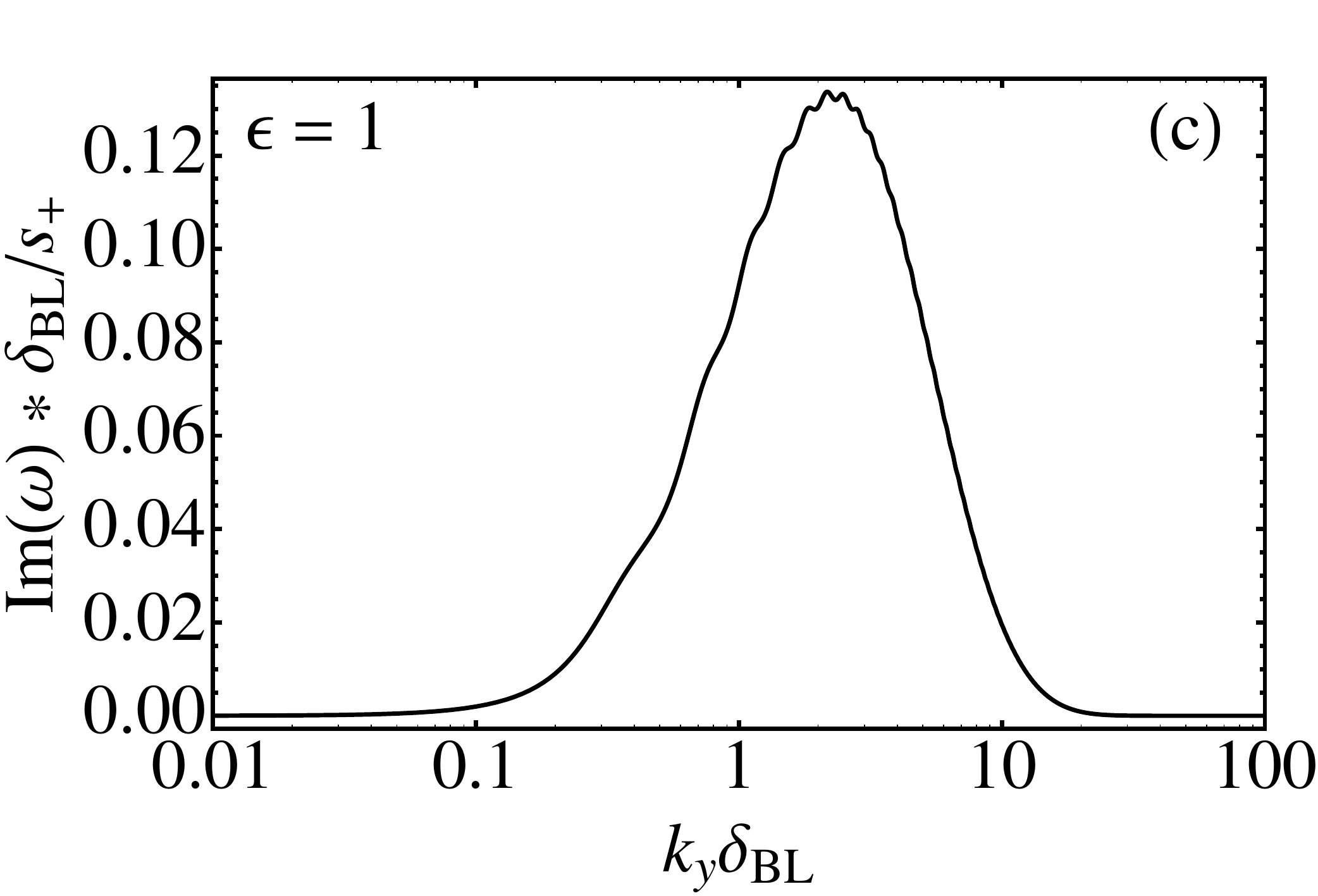}}
  \subfigure{\includegraphics[width=.32\textwidth]{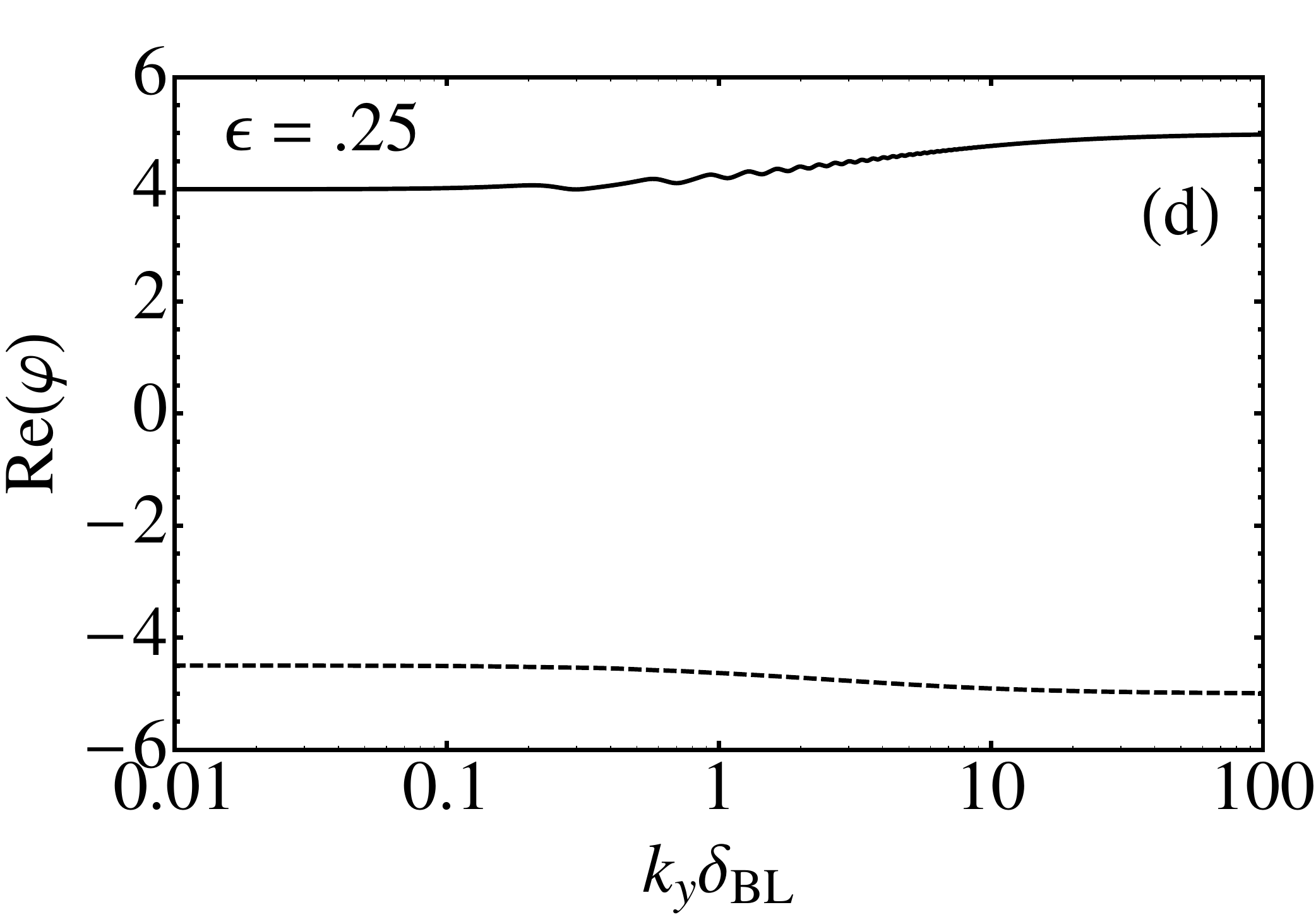}}
  \subfigure{\includegraphics[width=.32\textwidth]{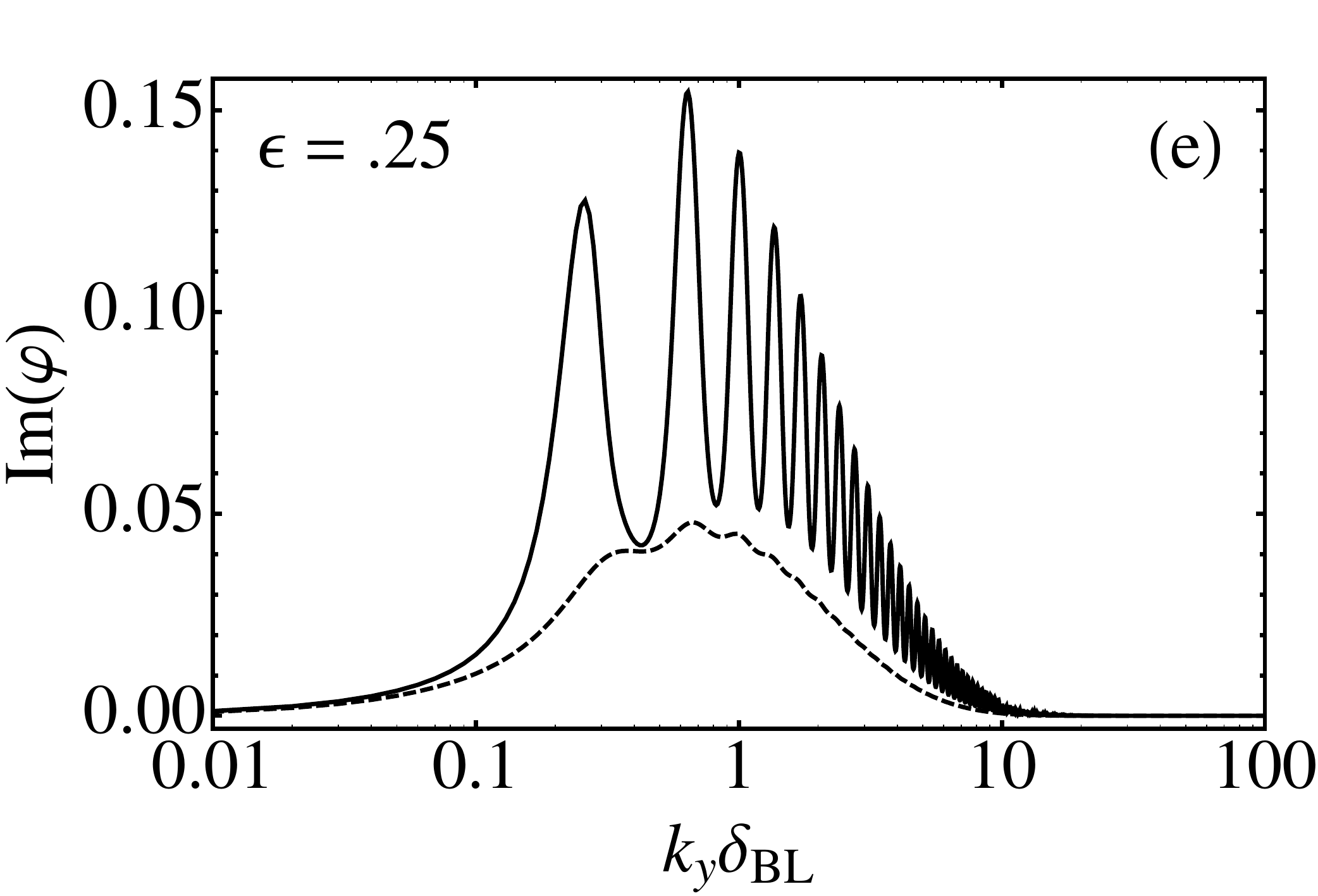}}
  \subfigure{\includegraphics[width=.32\textwidth]{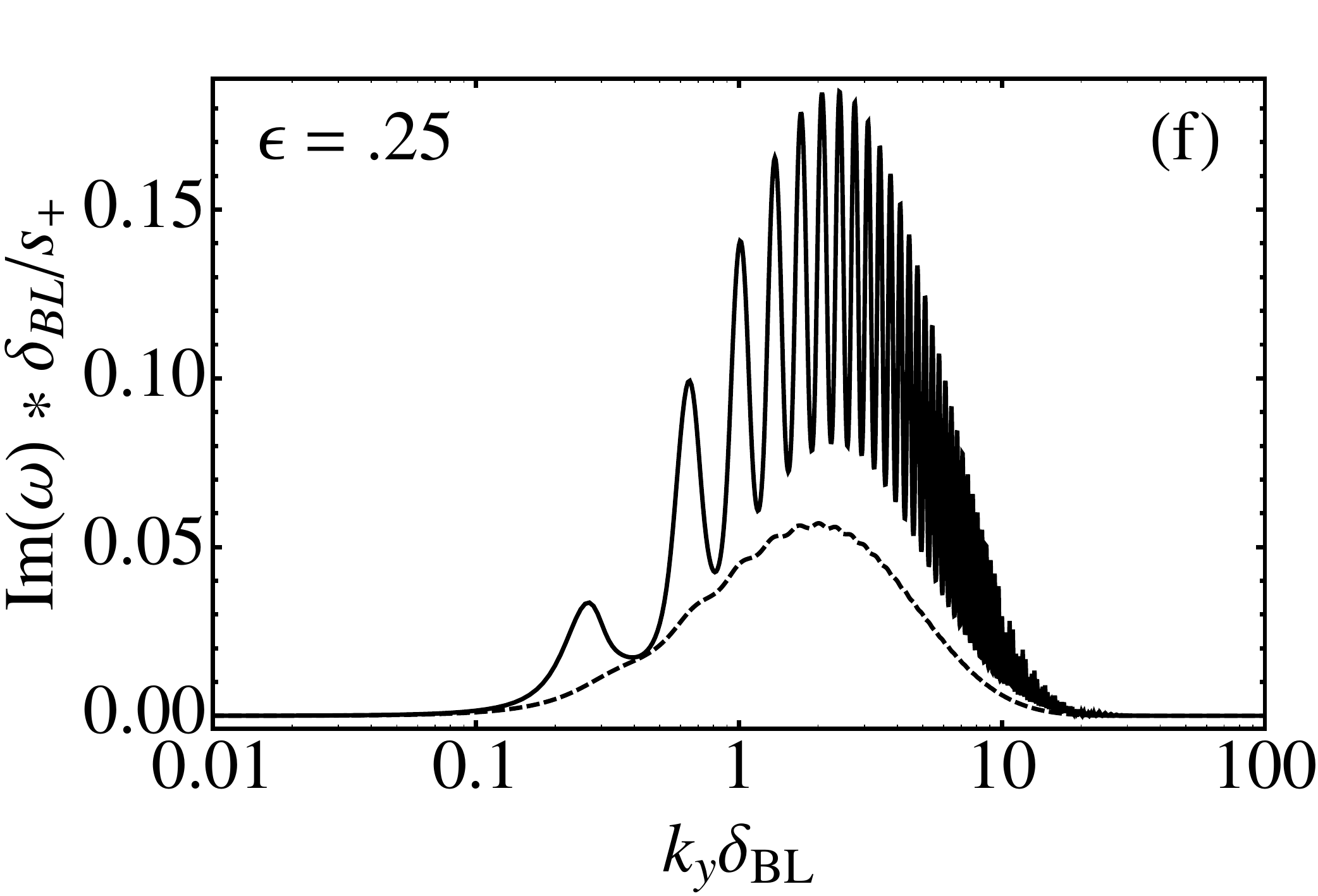}}
  \subfigure{\includegraphics[width=.32\textwidth]{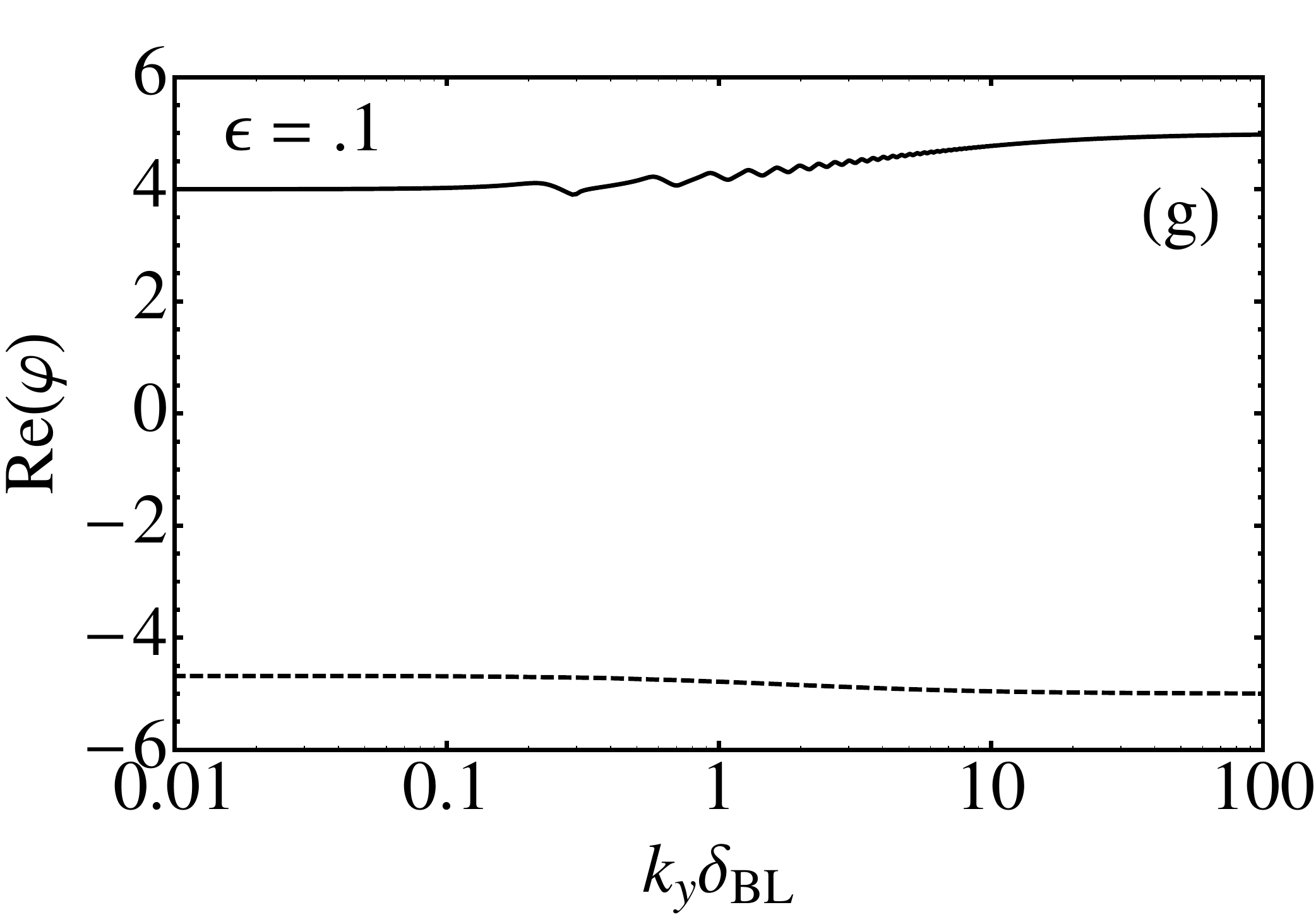}}
  \subfigure{\includegraphics[width=.32\textwidth]{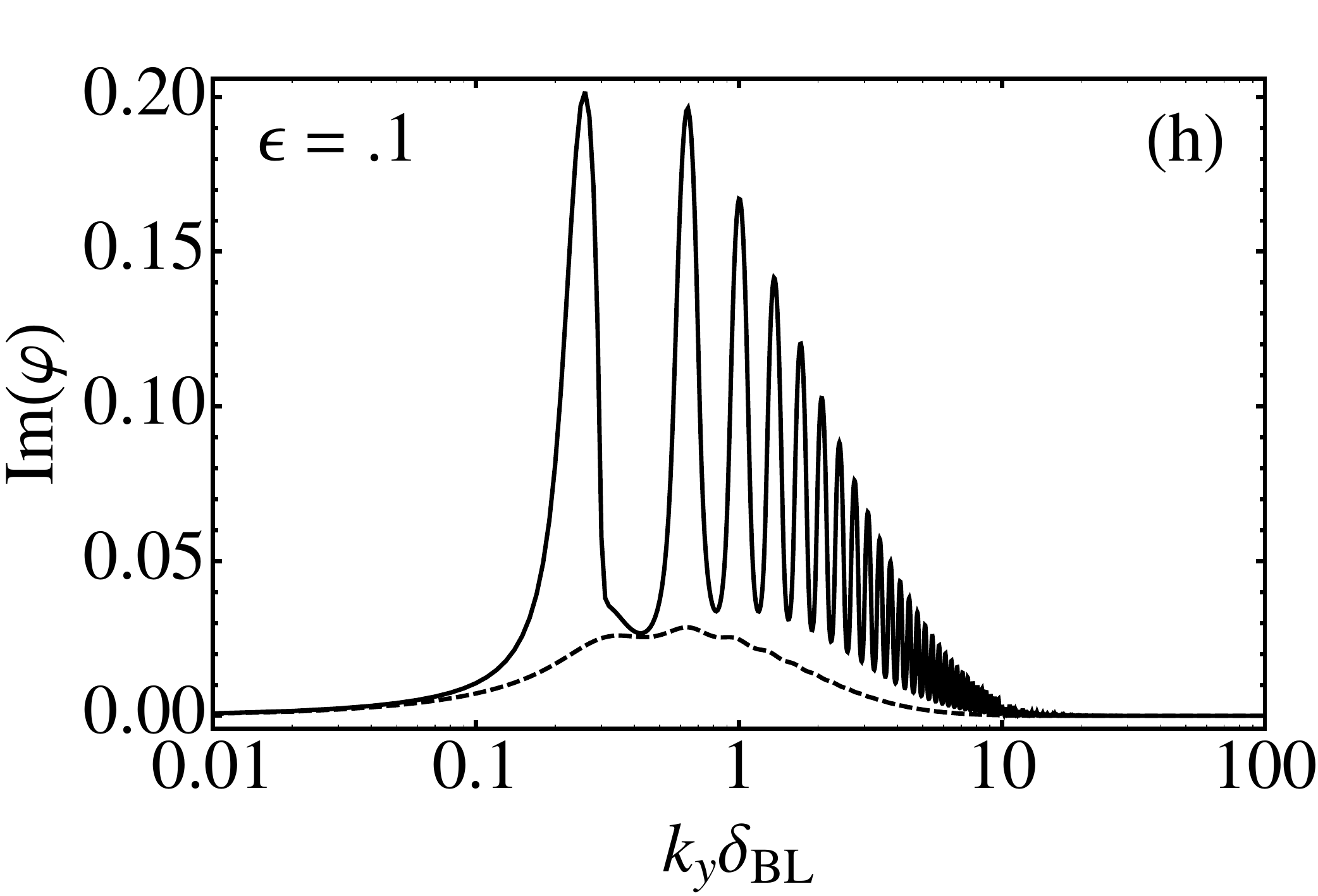}}
  \subfigure{\includegraphics[width=.32\textwidth]{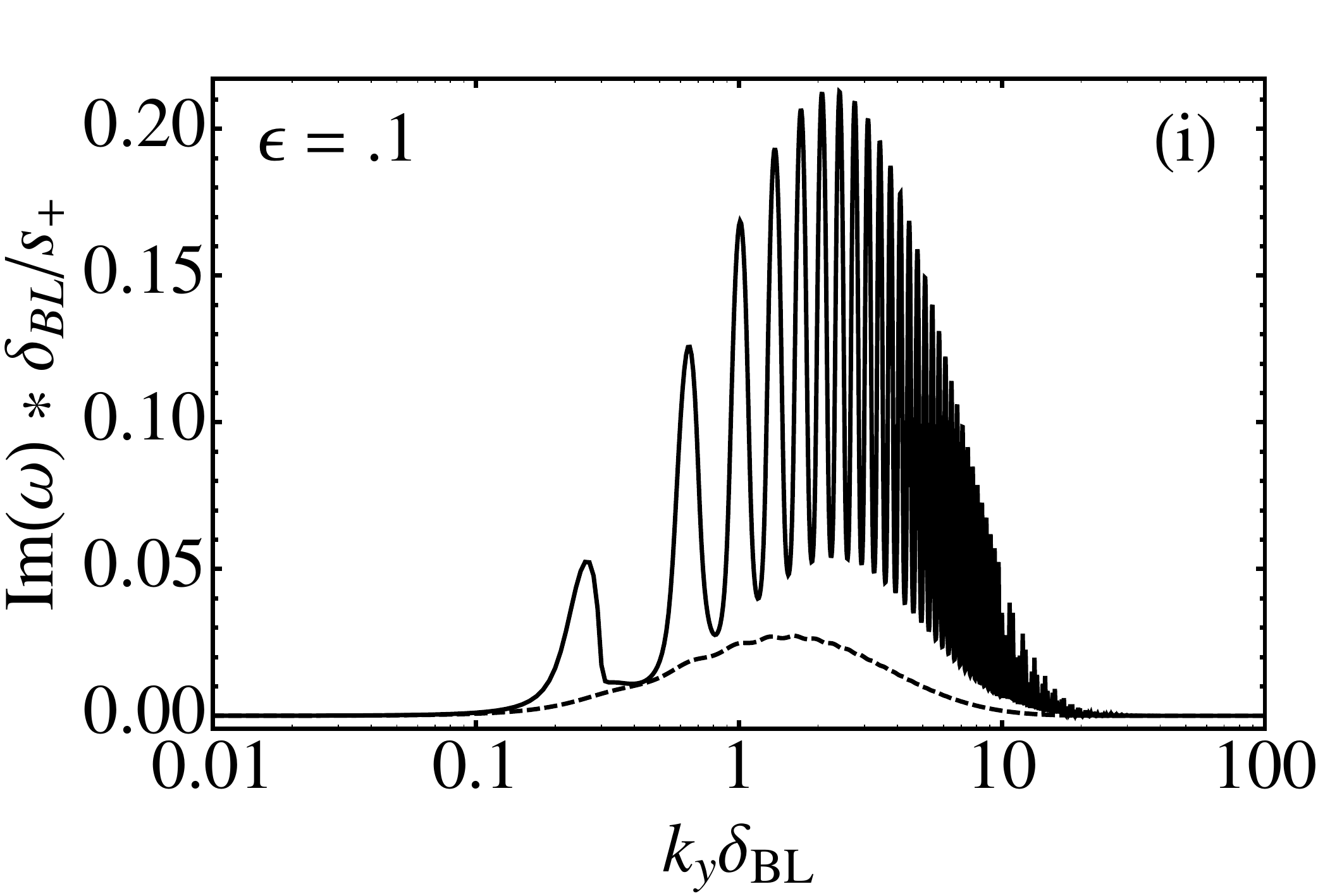}}
  \subfigure{\includegraphics[width=.32\textwidth]{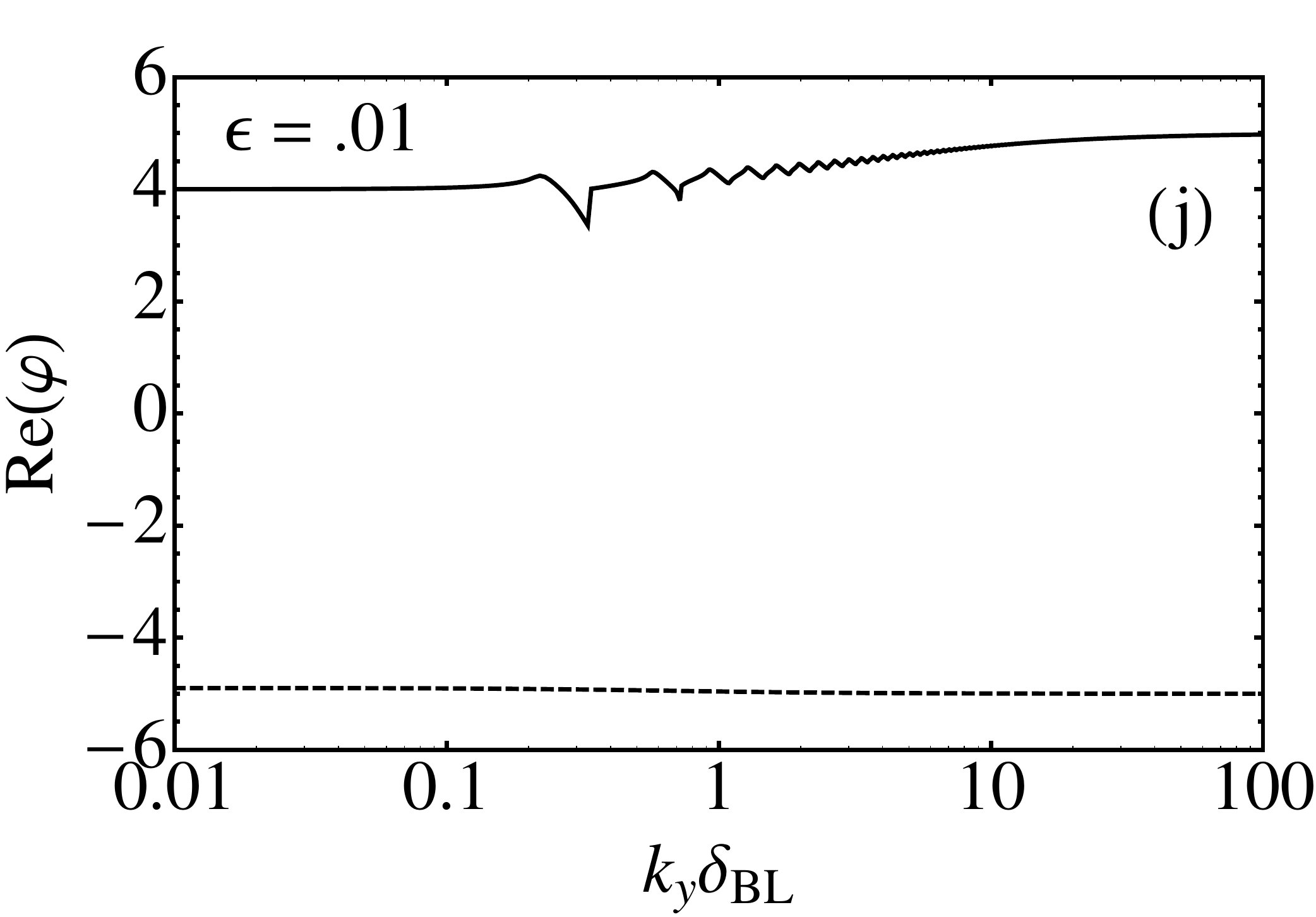}}
  \subfigure{\includegraphics[width=.32\textwidth]{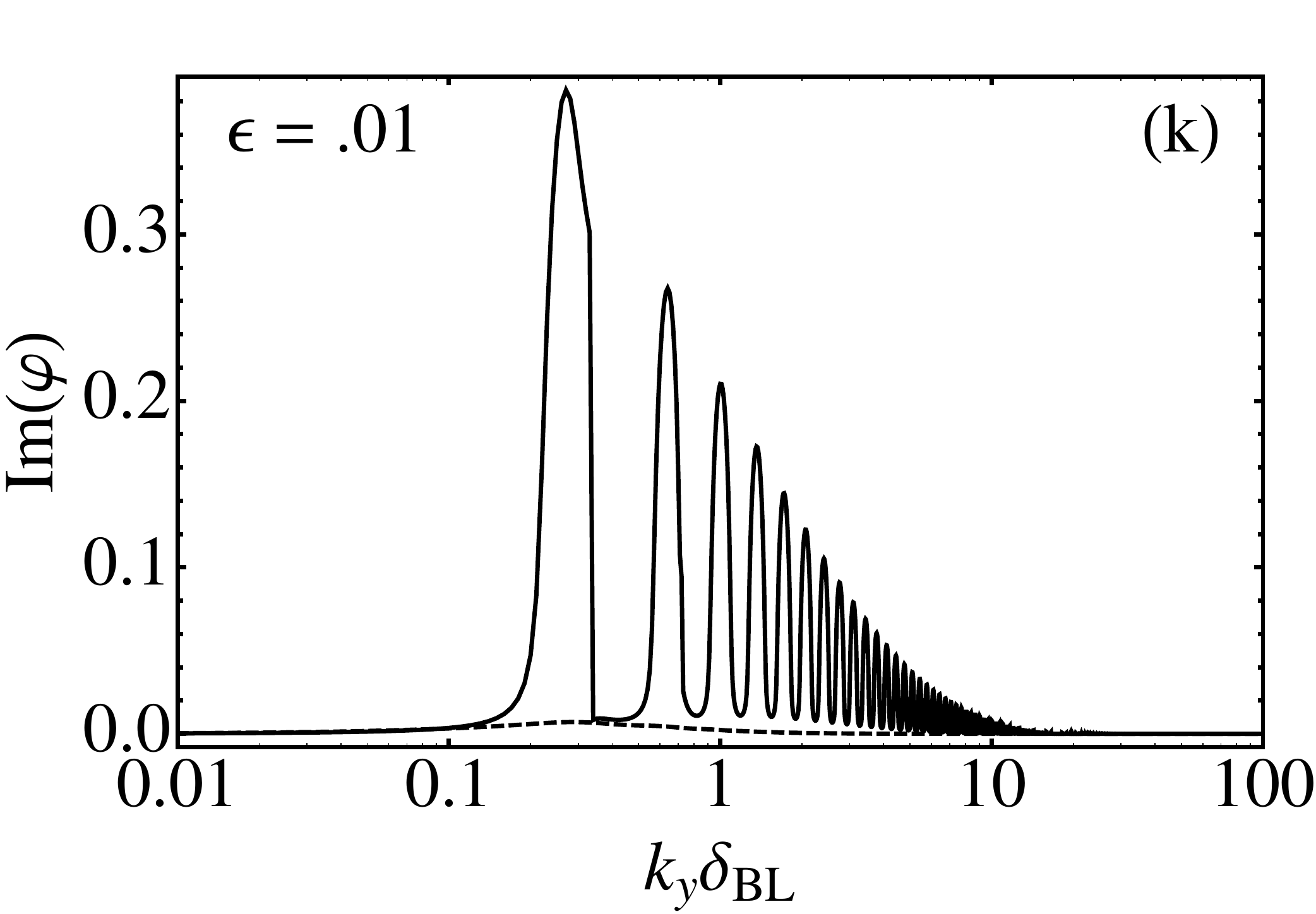}}
  \subfigure{\includegraphics[width=.32\textwidth]{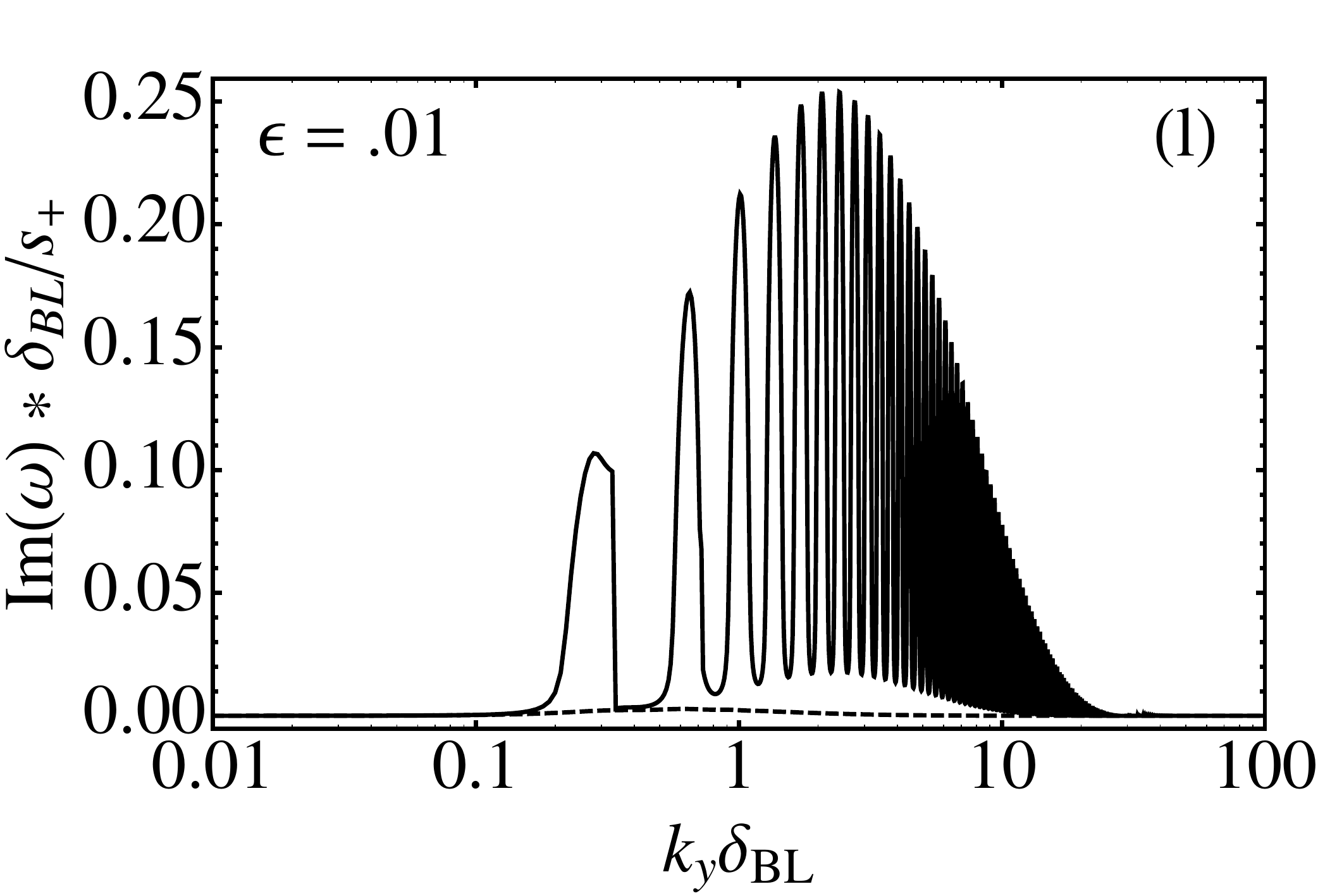}} 
  \subfigure{\includegraphics[width=.32\textwidth]{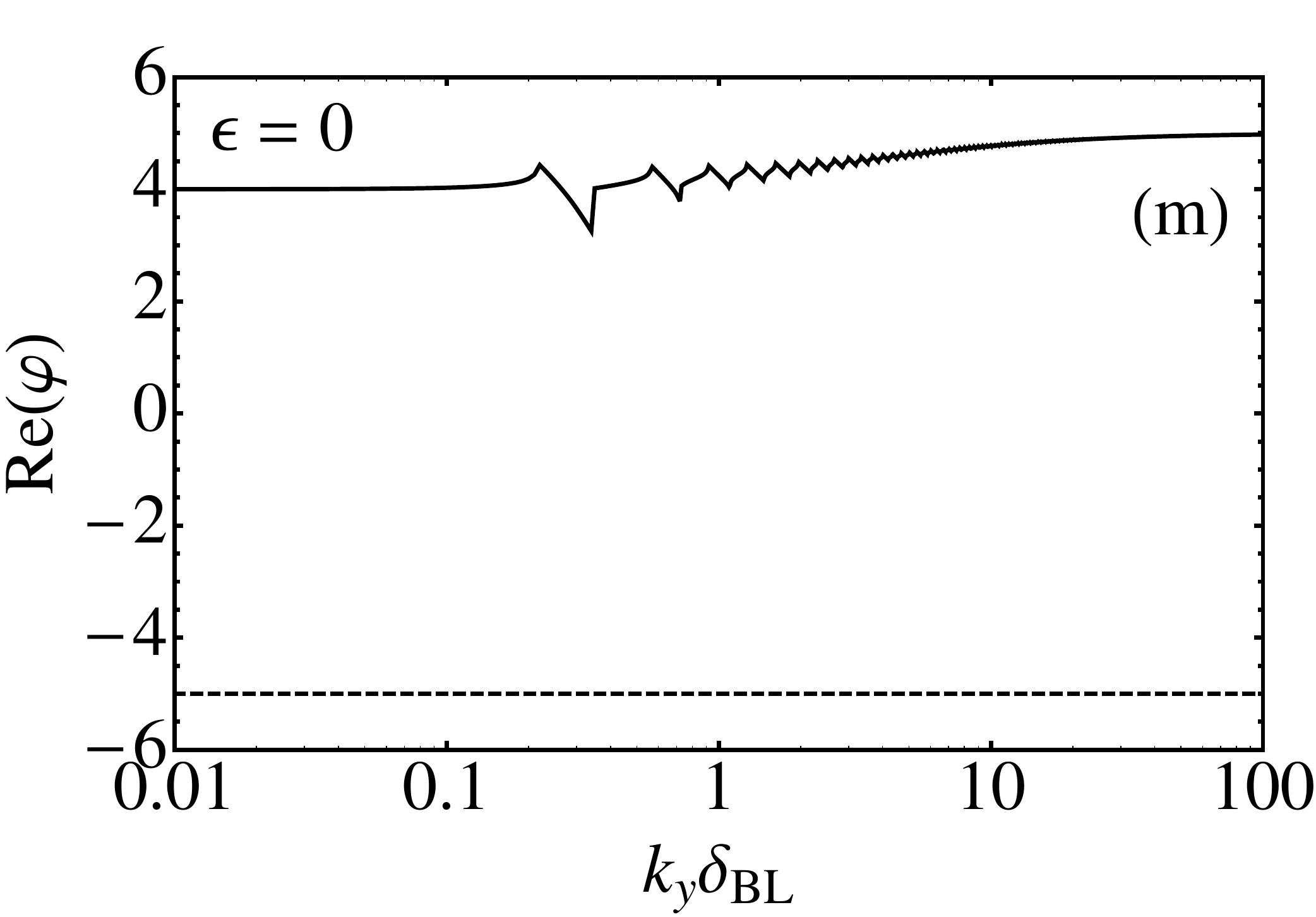}}
  \subfigure{\includegraphics[width=.32\textwidth]{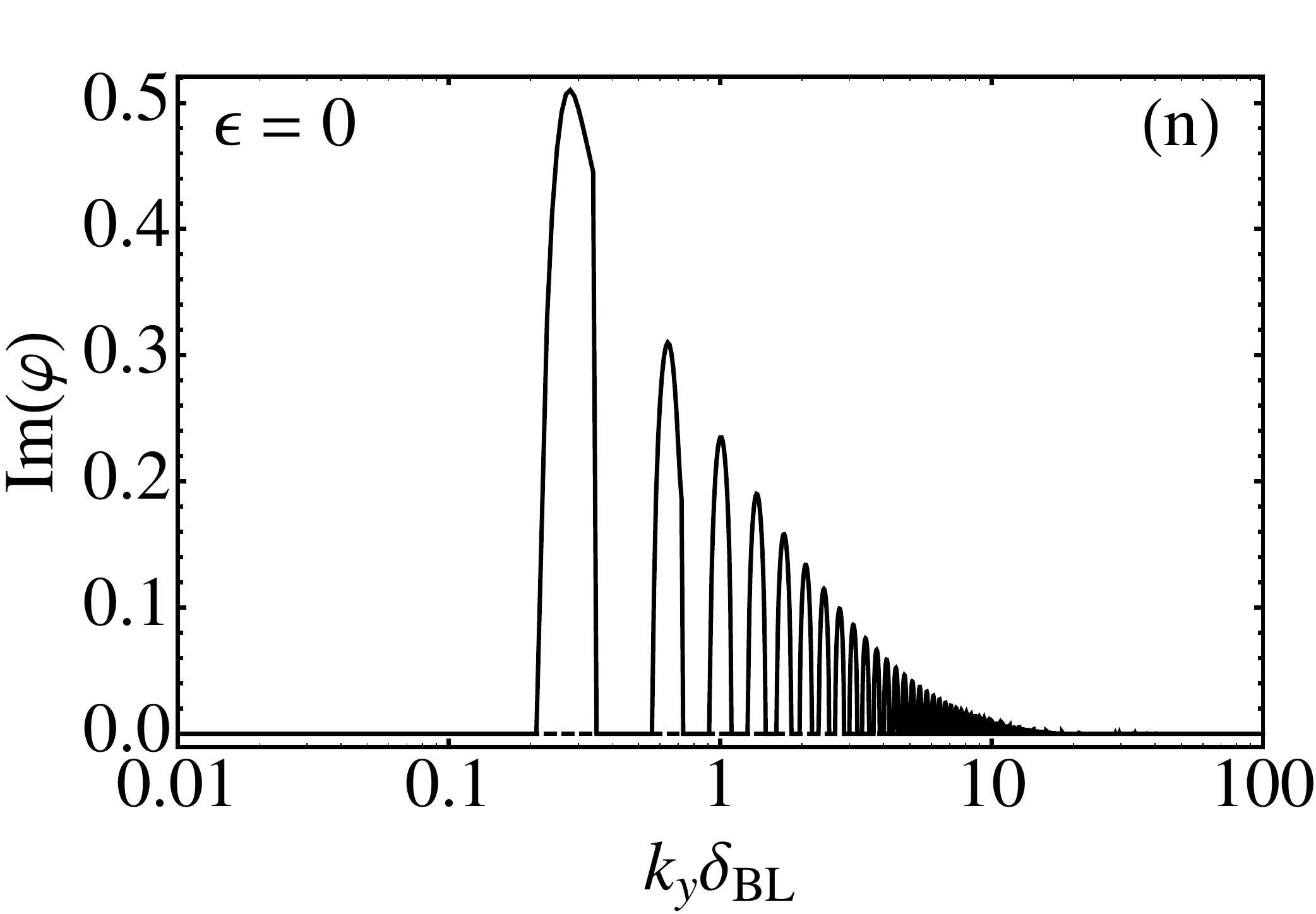}}
  \subfigure{\includegraphics[width=.32\textwidth]{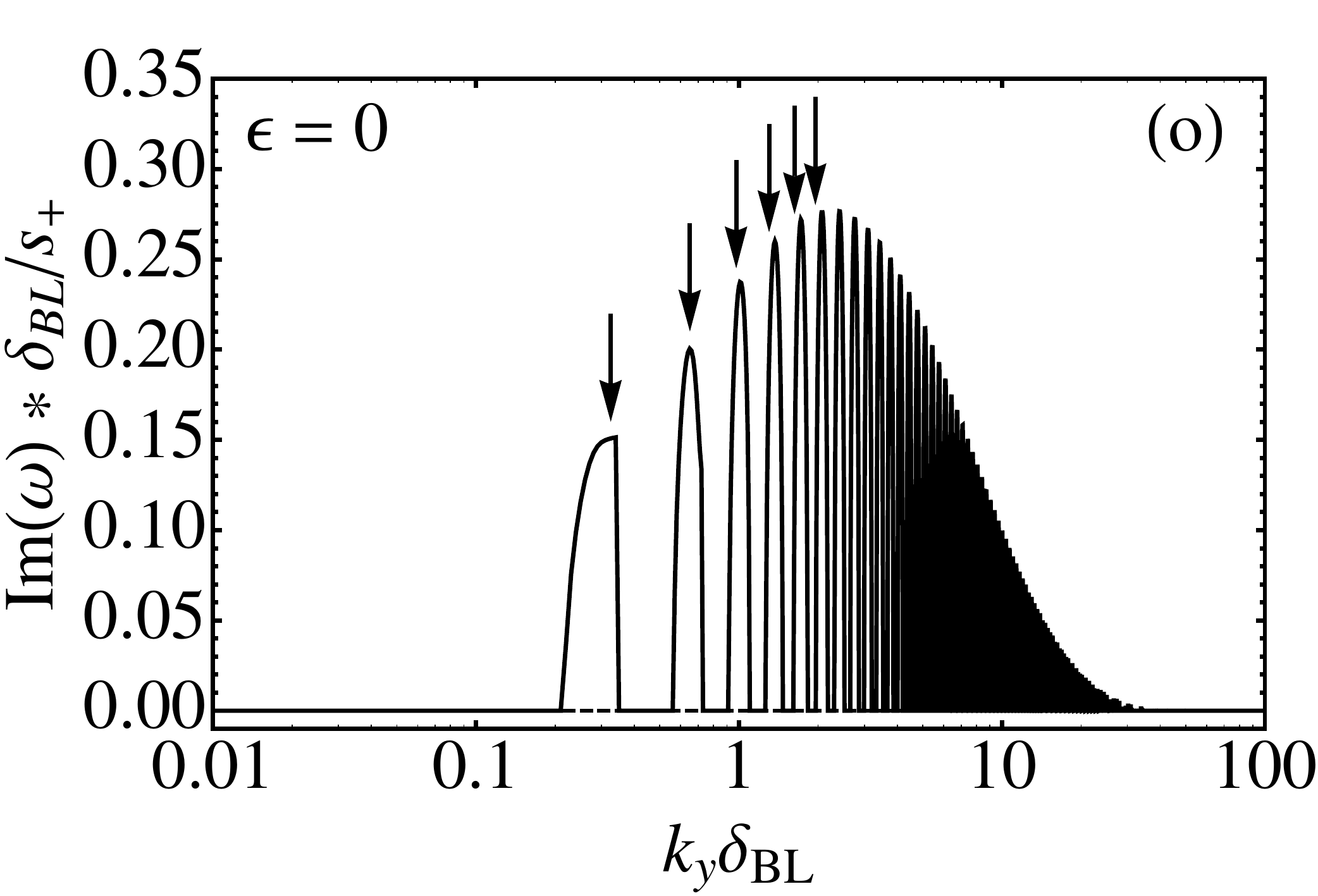}}   
  \caption{Dispersion relations for $M=5$ and a finite width shear
  layer. The top row is for $\eps = 1$, the second row for
  $\eps = .25$, the third row for $\eps = .1$, the fourth row for $\eps
  = .01$, and the bottom row for $\eps = 0$. The first column
  shows the real part of $\varphi$, the second the imaginary part of
  $\varphi$ and the third the imaginary part of $\omega$. The solid
  line corresponds to the upper branch and the dashed line to the lower
  branch. In the top row, the upper and lower branches have the same
  growth rate. The arrows in panel (o) show the wavenumbers
  corresponding to trapped wavemodes (\S \ref{physicalintuition}).}
\label{growthfinite}
\end{figure}

In Figure \ref{growthfinite}, we plot the
dispersion relation (\ref{dis_hyp}) as a function of wavenumber
for the parameters $M=5$ and $\eps = 1$, $\eps = .25$, $\eps =
.1$, $\eps = .01$, and $\eps = 0$. The $\eps = 0$ case is equivalent
to having a hard reflecting boundary at $x = -\delta_{BL}$. The modes
depicted are the ones that converge to the upper and
lower branches from \S \ref{compress_KH} in the vortex sheet limit
($k_y\delta_{BL} \ll 1$). For $\eps
< 1$, the upper branch always has a larger growth rate than the lower
branch, and both the upper and lower branches have the same growth rate
for $\eps = 1$. In the $\eps = 1$ case, we can identify the upper and
lower branches as the $n_\pm = 0$ decoupled modes in \S 5.4 of
\citet{Glatzel}. In addition to the upper and lower waves,
\citet{Glatzel} has shown that there is an infinite spectrum of damped
modes, but we do not consider these here, since we are interested in
determining the growth rate of the fastest growing mode. 

We now revisit the seemingly paradoxical
statement that the vortex sheet is stable to two dimensional
disturbances along the flow direction above a critical Mach number;
yet at the same time, the growth rate of the
fastest growing mode scales with $\delta_{BL}^{-1}$, which implies that the
thinner the shear
layer, the faster the instability proceeds. The key to resolving this
apparent controversy is to consider a mode having
$k_y\delta_{BL} \ll 1$. Its growth rate is diminished if we decrease
$\delta_{BL}$ while keeping $k_y$ constant for $M>M_\text{crit}$ and becomes
vanishingly small if we take the limit $\delta_{BL} \rightarrow 0$. This can be
seen from the dying left hand tail of the curve in
Figure \ref{growthfinite}b, and
keeping $k_y$ constant, while decreasing $\delta_{BL}$ we move
leftward along the
tail. This means $\Im[\omega]$ becomes smaller, since $\varphi = \omega/k_ys_+$
is directly proportional to $\omega$ for constant $k_y$ even as we
decrease $\delta_{BL}$.  We next point
out that the curve in Figure \ref{growthfinite}c remains unchanged in
shape or amplitude as we decrease
$\delta_{BL}$, keeping $k_y\delta_{BL}$ constant. Consequently, the value of $k_y$ for which the
maximum in $\Im[\omega]$ occurs $k_{y,\text{max}} \propto \delta_{BL}^{-1}$ and likewise
$\max[\Im[\omega]] \propto \delta_{BL}^{-1}$. Thus, as $\delta_{BL}$ is decreased, the
instability shifts to shorter wavelengths
and becomes more rapid, while at the same time, modes
having $k_y\delta_{BL} \ll 1$ are stabilized for a given value of $k_y$. Since any
real shear layer is likely to have
a nonzero width, one may therefore remark that taking
the vortex sheet limit for $M \gg M_\text{crit}$ masks the instability.

\begin{comment}
We can compare the dispersion relation for the $M \gg 1$ case to the $M \ll 1$
(incompressible) case for the setup given by Equations (\ref{vsetup}) and
(\ref{rhosetup}). For $M \ll 1$, the
dispersion relation is well known and is given by
e.g. \citep{ShtemlerMond}\footnote{Be wary of typos.} as
\ba
&&\left[\frac{\omega}{k}(1+\eps) - \eps \frac{\bar{V}}{k\delta_{BL}}\right] \times
\left[\frac{\omega}{k} - \frac{\bar{V}}{k\delta_{BL}}F(2k\delta_{BL})\right] = -\eps
\left(\frac{\omega}{k}\right) \left(\frac{\bar{V}}{k
  L}\right)\left[G(2kL) + 1\right], \\
&&F(x) = \frac{1}{2}\exp(-2x) - \frac{1}{2} + x, \ G(x) = \exp(-2x) -
1 \\
&&F(x) = x^2 + \mathcal{O}(x^3), \ G(x) = -2x + \mathcal{O}(x^2) \text{
  for } x \ll 1.
\ea
Just as for $M \gg 1$, the fastest growing mode again has $k_yL
\sim 1$, $k_{y,\text{max}} \propto L^{-1}$, and $\max[\Im[\omega]] \propto
L^{-1}$. Now, taking the limit $L \rightarrow 0$, while
keeping $k_y$ constant, the instability growth rate converges to 
a non-zero value, and we recover the conventional KH dispersion 
relation in the vortex sheet limit (Equation (\ref{KHwellknown})).
\end{comment}

%%%%%%%%%%%%%%%%%%%%%%%%%%%%%%%%%%%%%%%%%%%%%%%%%%%%%

\subsection{Numerical Verification}

\begin{figure}[!h]
  \centering
  \subfigure[]{\includegraphics[width=.6\textwidth]{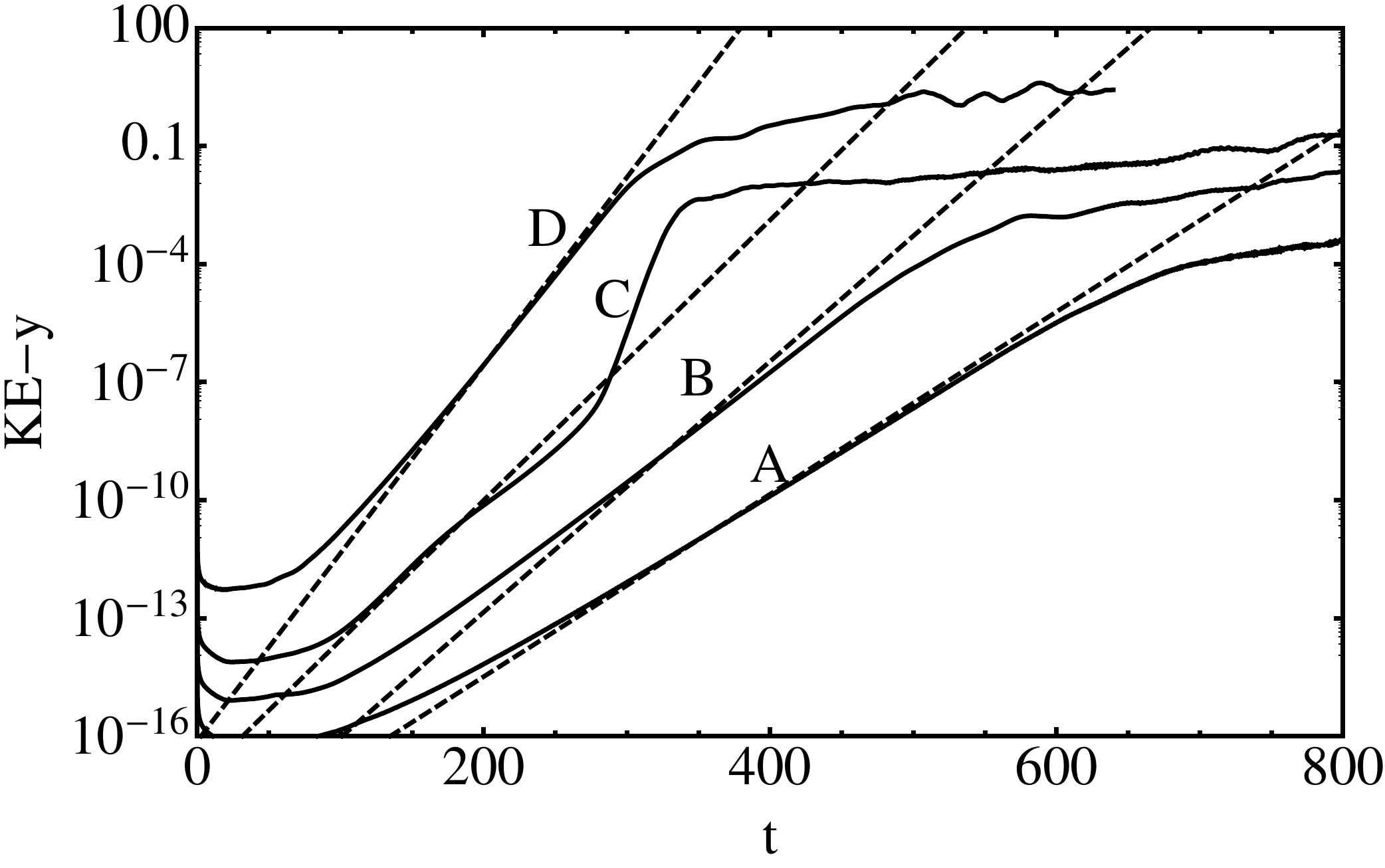}}  
  \caption{Comparison of the analytically derived growth rates (given
  by the slopes of the dashed
  lines) to the ones obtained using Athena (solid lines). The curves A, B,
  C, D correspond to $\eps = 1$, $.25$, $.1$, and $0$, respectively
  (see text for
  further simulation details) and have been offset vertically for
  clarity.
  }
\label{athenatest}
\end{figure}

To independently verify our dispersion relation (\ref{dis_hyp}), we
ran direct hydrodynamical simulations of shear layers using
the Godunov code Athena \citep{Stoneetal} and compared the growth rate
of the fastest growing mode predicted by our dispersion relation to that
obtained in the simulations. We initialized a flow along the $y$-direction
using the setup described by Equations (\ref{vsetup}), and for all of
the simulations we used $\gamma = 5/3$, $\delta_{BL}=1$ for the half width of
the shear layer, and periodic boundary
conditions in the $y$-direction; Table \ref{flowpartable}
summarizes the simulation specific parameters. To seed the
instability, we initialized random perturbations to $v_x$ of magnitude
$10^{-6}$ in the region $x > -1$. We found that for the $M=5$ runs, we
needed a high resolution in the $x$-direction to get converged estimates for the
growth rates. Figure \ref{athenatest} shows $\rho v_x^2/2$ averaged over box as a
function of time. The
solid lines show simulation results, and the dashed lines
show the predictions from considering the fastest growing mode for the
upper branch, using the dispersion relation (\ref{dis_hyp}). There is
good agreement between the two, especially for the limiting cases of
$\eps = 0$ and $\eps = 1$.

\begin{table}[!h]
\begin{center}
\begin{tabular}{|c|c|c|c|c|c|c|}
\hline
Label & $M$ & $\eps$ & $x$-range & $y$-range & $N_x \times
N_y$ & (BC-$x1$, BC-$x2$) \\
\hline
A & 5 & 1 & (-4,4) & (-16,16) & $8192 \times 8192$ & (outflow, outflow) \\
B & 5 & .25 & (-4,4) & (-16,16) & $8192 \times 8192$ & (outflow, outflow) \\
C & 5 & .1 & (-4,4) & (-16,16) & $8192 \times 8192$ & (outflow, outflow) \\
D & 5 & 0 & (-1,4) & (-16,16) & $2048 \times 2048$ & (reflecting, outflow) \\
E & .05 & 1 & (-4,4) & (-8,8) & $512 \times 512$ & (reflecting, reflecting) \\
\hline
\end{tabular}
\end{center}
\caption{Parameters for the simulations with Athena. $M$ -- Mach
  number, $\eps$ -- density ratio, $x,y$-range -- box size in
  $x,y$-direction, $N_x \times N_y$ -- number of cells in
  $x,y$-direction, (BC-$x1$, BC-$x2$) -- lower and upper boundary
  conditions in $x$ direction.}
\label{flowpartable}
\end{table}

\subsection{Physical Intuition}
\label{physicalintuition}
\begin{figure}[!h]
  \centering
  \subfigure[]{\includegraphics[width=.49\textwidth]{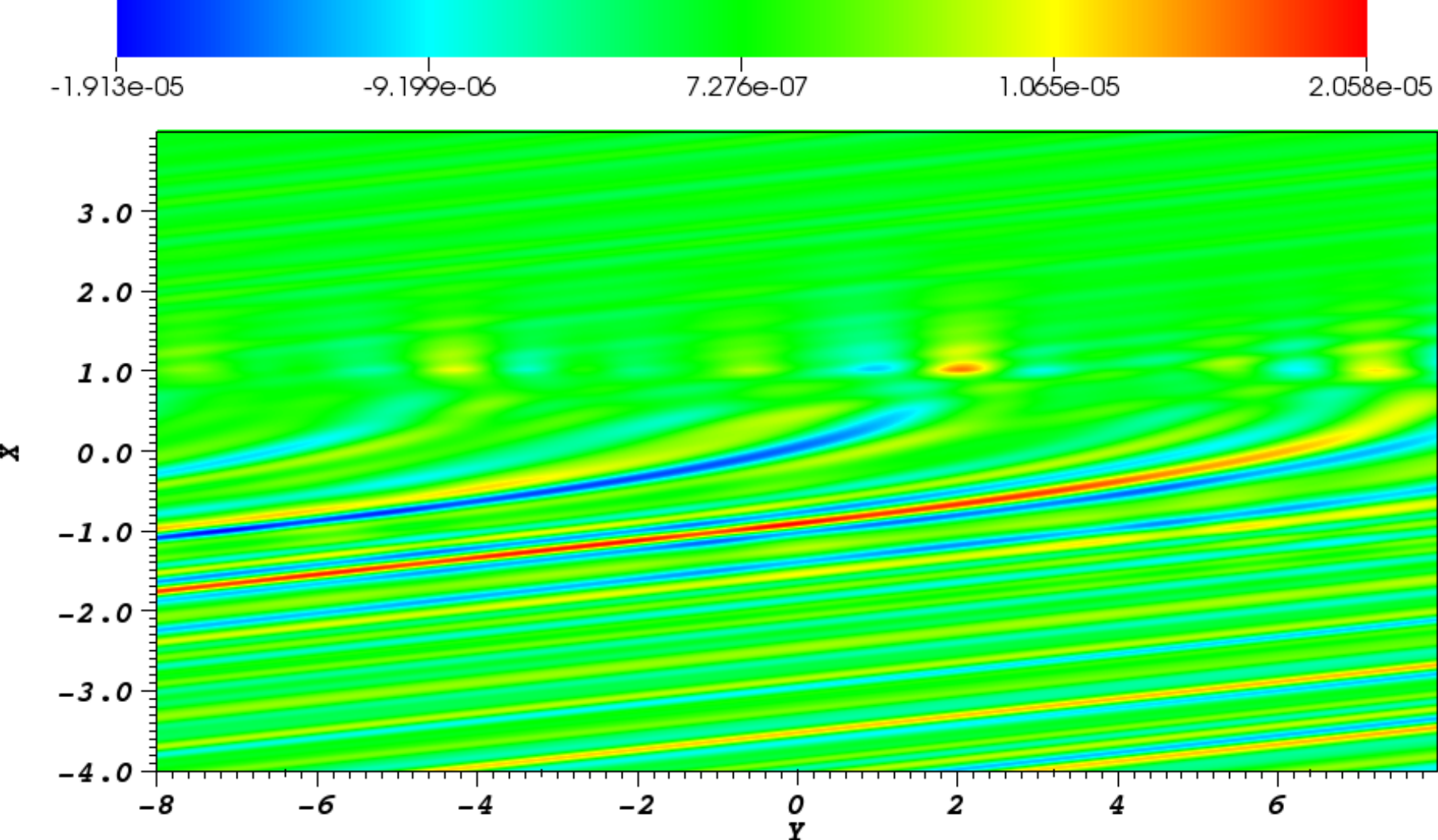}}
  \subfigure[]{\includegraphics[width=.49\textwidth]{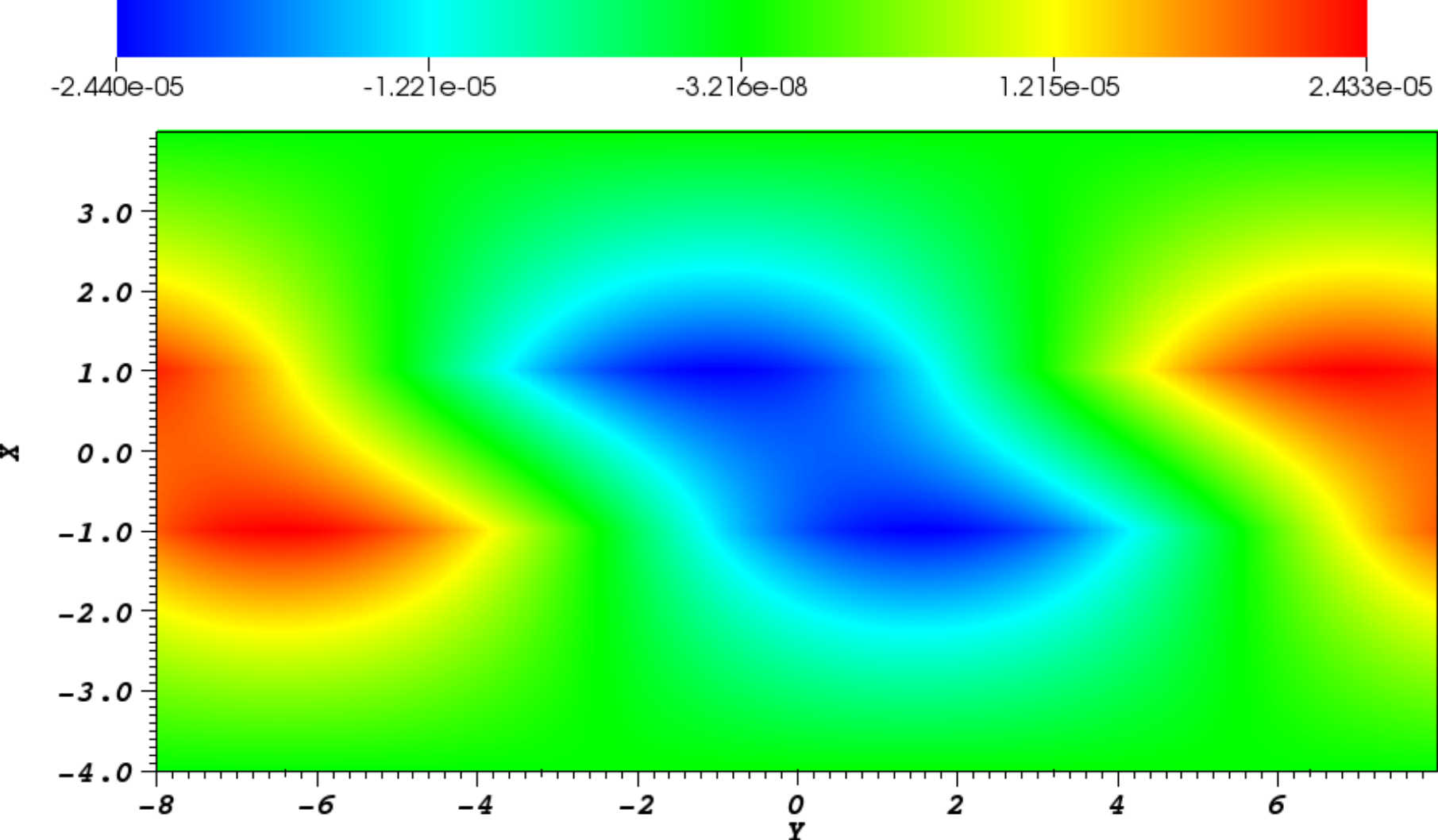}} 
  \subfigure[]{\includegraphics[width=.75\textwidth]{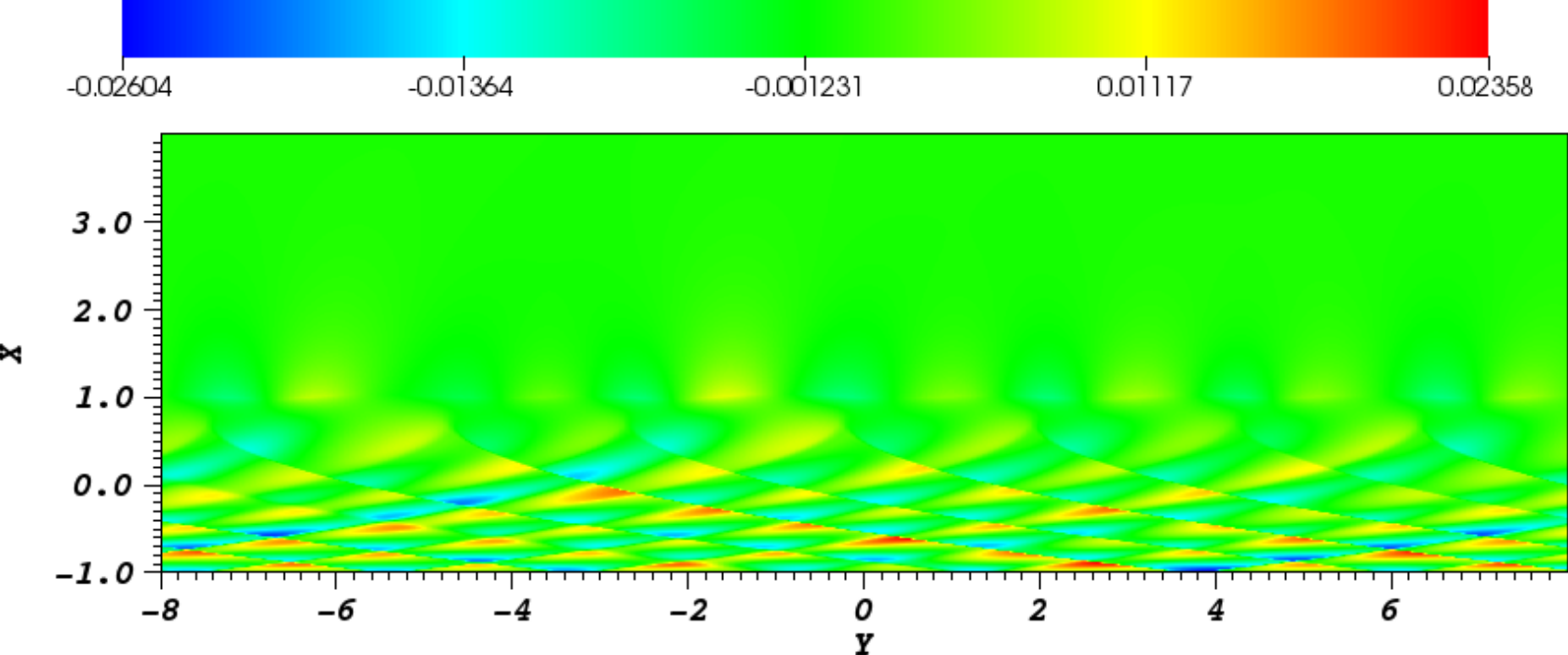}} 
  \caption{Panels (a), (b), and (c) show $v_x$ during the linear stage of the
  instability for simulations A, E, \& D respectively and have been
  rotated by 90 degrees.}
\label{eigenfig}
\end{figure}

Panels, (a), (b), and (c) of Figure \ref{eigenfig} show the spatial structure of 
$v_x$ for simulations A, E \& D during
the linear stage of the instability. The
fastest growing modes in these three cases are very different, illustrating the
different instability mechanisms which operate for the $M \gg M_\text{crit}$ vs. $M
\ll M_\text{crit}$ cases and also for $\eps = 1$ vs. $\eps=0$. The
case $M \ll M_\text{crit}$, in panel (b) has been extensively studied
(e.g. \citep{Vallis}), so we will not consider it here.. 

For the $M \gg M_\text{crit}$ case with $\eps = 1$, the instability is
  caused by a {\it radiation mechanism}. This mechanism has already
  been discussed by \citet{Glatzel}, so we only mention it here
  briefly. Each mode of the dispersion relation can be associated with
  a pseudo-energy that is conserved. If a mode has a negative
  pseudo-energy, then radiation of energy away from the boundary layer
  region will cause the pseudo-energy to become even more negative,
  amplifying the mode and leading to instability. The radiation
  mechanism is responsible for the smooth, broad hump in panels
  (b) and (c) of Figure \ref{growthfinite}. 

For the case of $M \gg M_\text{crit}$ and $\eps = 0$, the instability
  mechanism is quite different. Rather than a broad hump, panels (n)
  and (o) of Figure \ref{growthfinite} exhibit sharp localized
  peaks. The physical cause of these peaks is due to
  over-reflection of modes that become trapped between the lower
  boundary and the critical layer.
  Thus, we shall call this the {\it over-reflection mechanism}. 

  The over-reflection mechanism is discussed in \citet{NGG}, who
  considered a rotating shear layer adjacent to a reflecting
  wall. They explained the instability in the $\eps = 0$ case in terms
  of the leaking of
  action density past the corotation radius, which is the analog of the
  critical layer for a rotating system. Like the pseudo-energy of
  \citet{Glatzel}, the action density is a conserved
  quantity, and instability occurs because a
  wavemode that is trapped between the critical layer and the wall has
  a negative action density, whereas positive action leaks out past
  the critical layer. As a result, the amplitude of the trapped
  wavemode becomes even more negative and grows with time. Put in this way, it is
  clear that the over-reflection and radiation mechanisms are
  related. However, they lead to a very different structure for the
  dispersion relation, as evidenced by panels (n) and (o) vs. panels
  (b) and (c) of Figure \ref{growthfinite}. Thus, we prefer to regard them as
  separate mechanisms, but point out that in both cases, instability is
  ultimately caused by radiation emitted from the boundary layer region.

  We now elucidate some of the properties of the modes that are trapped between
  the critical layer and the lower reflecting boundary using a simple
  model. Consider a sonic mode that is trapped between the lower boundary and the
  critical layer. If the mode has a well-defined
  phase velocity, then by performing a velocity boost one can
  transform into a frame in which the wavefronts are stationary. Let us
  work in this frame, since it makes the explanations more clear. In
  order for the wavefronts to be stationary, the equation for a wavefront is
  \ba
  \label{wavefronteq}
  \frac{d y}{d x} = \pm \sqrt{\mathcal{M}(x)^2 - 1}.
  \ea 
  Here, $\mathcal{M}(x)$ is the Mach number as a function of $x$, and
  the positive sign is for waves propagating in the $-x$ direction, while
  the negative sign is for waves propagating in the $+x$
  direction. Equation (\ref{wavefronteq}) is obtained from the fact
  that the wavefront propagates at the speed of sound, and in order
  for it to be stationary, the angle
  that the wavefront forms with the $y$-direction obeys $\sin(\theta) =
  1/\mathcal{M}(x)$.  

  Consider now the upper branch from \S \ref{ratiomach}, for which the
  critical layer is at
  $x_{c} = \delta_{BL}(M-1)/M$. At the critical layer, we set $dy/dx =
  0$, which is just to say that an upward traveling sound wave is
  reflected there. It immediately follows that inside the region of shear
  \ba
  \label{Mxeq}
  \mathcal{M}(x) =  -M\left(1-\frac{x}{\delta_{BL}}\right), \ -\delta_{BL} \le x \le \delta_{BL}. \\
  \ea
  Thus, the frame in which sound waves are stationary and reflect at
  the critical layer is boosted by $-M$ relative to the
  frame we defined in Equations (\ref{vsetup}).

\begin{figure}[!h]
  \centering
  \subfigure[]{\includegraphics[width=.49\textwidth]{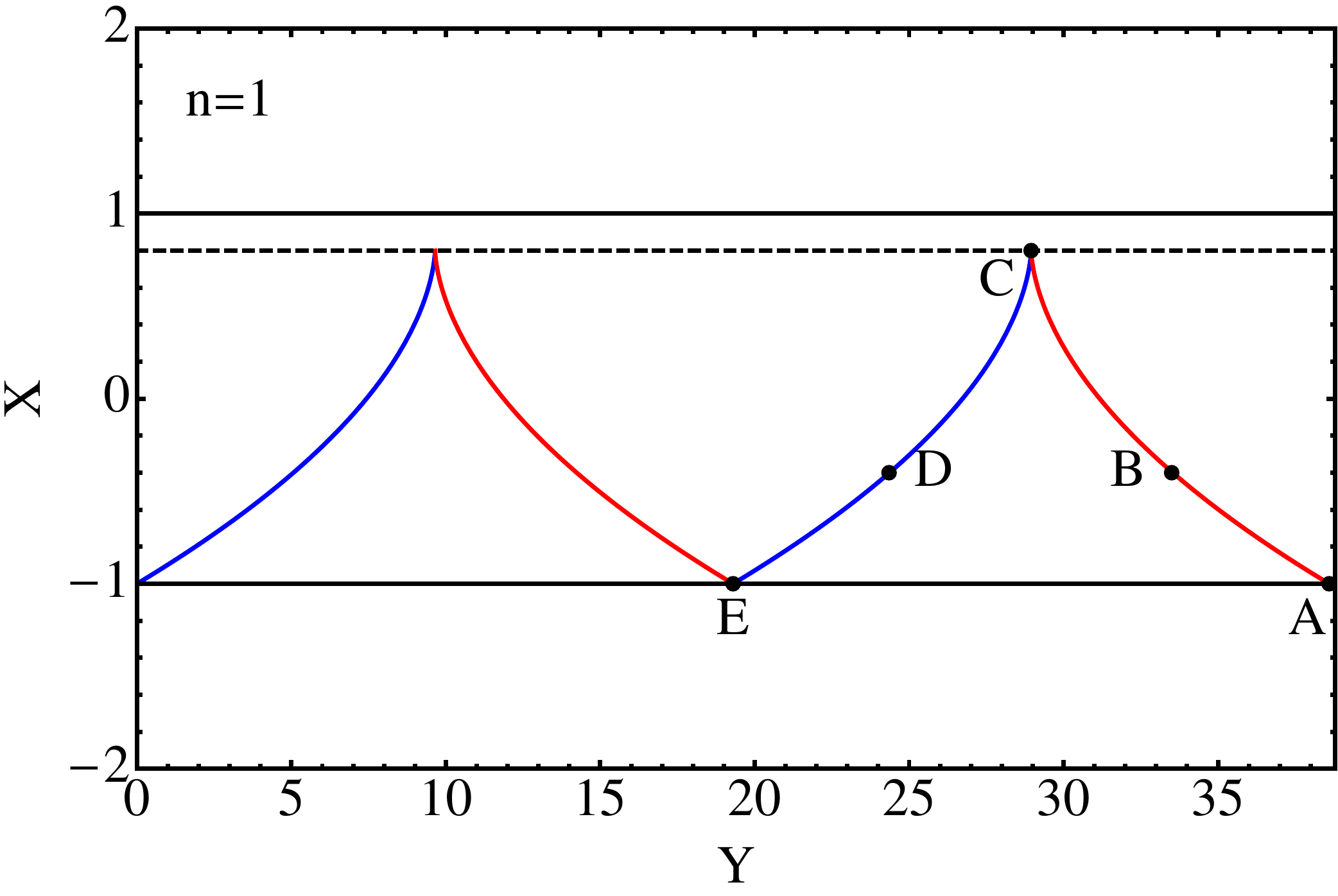}}
  \subfigure[]{\includegraphics[width=.49\textwidth]{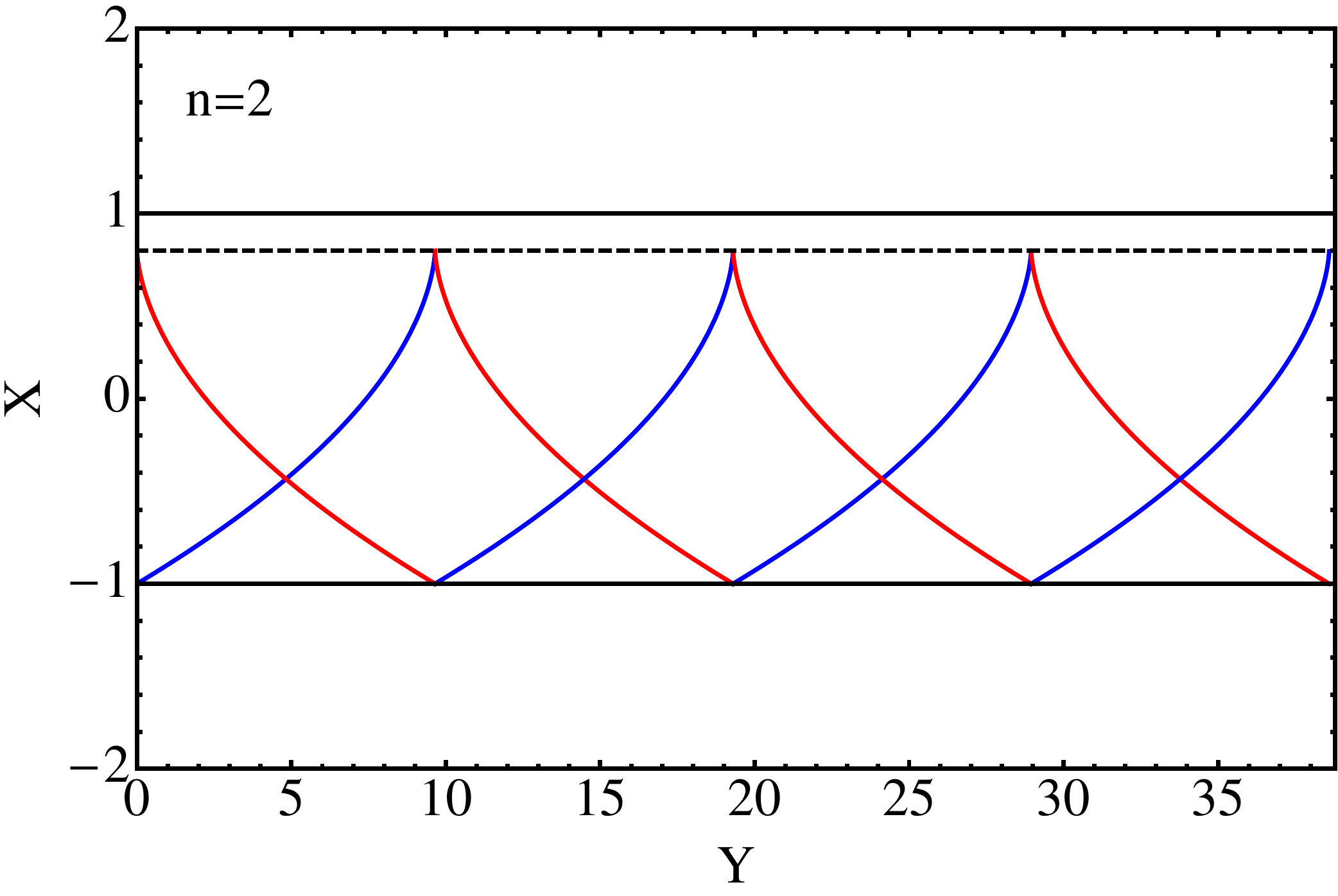}} 
  \caption{The $n=1$ and $n=2$ trapped modes for $M=5$ and
  $\delta_{BL} = 1$. The lower black line denotes the solid
  lower boundary, the upper black line denotes the top of the shear
  layer, and the dashed black line denotes the critical layer. The red
  segments correspond to propagation in the $+x$ direction and the
  blue segments correspond to propagation in the $-x$
  direction.}
\label{wavefrontfig}
\end{figure}

  We now consider the fate of a wavepacket that is trapped between the lower
  boundary and the critical layer. Figure \ref{wavefrontfig} shows a
  set of wavefronts derived by integrating Equation
  (\ref{wavefronteq}) with red segments corresponding to propagation in
  the $+x$ direction and blue segments corresponding to propagation
  in the $-x$ direction. Consider a localized wavepacket at point A in
  Figure \ref{wavefrontfig}a. The wavevector of the wavepacket is
  oriented perpendicular to the wavefront, and as the wavepacket
  propagates towards point B, its wavevector is rotated by the shear.
  When the wavepacket reaches the critical layer at point C, it is
  reflected back towards point D, and the reflected wavepacket has a
  higher amplitude than the incident one (see \citet{NGG} for the case
  of a rotating shear layer). After
  reflecting off the critical layer, the wavepacket propagates
  downward to point D, its wavevector continuing to be
  rotated in the same sense as before due to the shear. Finally the
  wavevector reaches the perfectly
  reflecting boundary at point E, whereupon the $x$-component of its
  wavevector is reflected and the cycle begins anew. It is thus clear
  that the repeating cycle A-E will lead to exponential amplification
  of the wavepacket, due to over-reflection at the
  critical layer.

  The distance in $y$ between points A and E is the wavelength of the
  longest wavelength mode, $\lambda_{\text{max},y}$,
  that can be trapped between the lower boundary and the critical
  layer. A trapped mode can have a wavelength shorter than
  $\lambda_{\text{max},y}$, as long as its wavelength satisfies the
  relation $\lambda_y = \lambda_{\text{max},y}/n$ where $n$ is an
  integer. Thus, the relation for the wavenumber of the $n$-th trapped mode in
  the general case is 
  \ba
  k_{n,y} = \frac{2 \pi n}{\lambda_\text{max,y}}.
  \ea
  We indicate using arrows the values of $k_{n,y}$ for the first six
  trapped modes in panel (o) of Figure \ref{growthfinite}, and it is
  clear that they agree well with the locations of the peaks in the
  dispersion relation, which lends support for the
  over-reflection argument. We also plot the $n=2$ case in
  Figure \ref{wavefrontfig}b and point out the similarity between the
  analytical curves in Figure \ref{wavefrontfig} and the shape of the wavefronts from
  simulation D in Figure \ref{eigenfig}c.

  We have discussed the radiation mechanism for $\eps=1$ and
  the over-reflection mechanism for $\eps=0$. We now discuss what
  happens in the more general case of $0 < \eps < 1$. As we can see
  from Figure \ref{growthfinite}, {\it both} mechanisms operate
  simultaneously for $0 < \eps < 1$. In this case, there is partial
  reflection at the lower boundary, and as $\eps$ goes from one to zero, less
  and less of the radiation can escape from the boundary layer region,
  so the radiation mechanism becomes weaker and weaker. This is
  evidenced by the decreasing size of the bump in the second and third
  columns of
  Figure \ref{growthfinite} as $\eps$ goes to zero. For $\eps = 0$, the
  bump disappears entirely, and the radiation mechanism no longer
  operates. On the other hand, as $\eps$ goes from one to zero the reflection
  mechanism becomes stronger and stronger, since more and more of the
  energy is reflected at the lower boundary with total reflection at
  $\eps = 0$. Interestingly, there are still some small-scale wiggles in the
  dispersion relation, even for $\eps = 1$. This is because
  even if the density is everywhere uniform, radiation can still partially reflect off the
  discontinuity in the velocity derivative at $x = -1$ \citep{Glatzel}. However,
  reflections off a discontinuity in the velocity derivative are quite weak, so
  the wiggles are small, and the radiation mechanism is dominant.

\section{Discussion and Conclusions}
\label{discon}

We have studied supersonic shear instabilities that could drive the
turbulence in the BLs of stars for which the disk is undisrupted by a
magnetic field. Our study is aimed mainly at identifying the
  instabilities that lead to the formation of the BL when the disk
  just touches the surface of the star. The
main result of our work is the identification of two types of
instabilities that could operate in the BLs of such systems and had
not been previously discussed in this context. 

The first is an
instability of a vortex
sheet at high Mach number caused by gravity. Although the vortex sheet
is stable above a critical Mach number, the addition of a small amount
of gravity destabilizes it. We have found that the eigenfrequencies of
the dispersion relation in the limit $G \rightarrow 0$ acquire a purely
imaginary term, which is proportional to the small parameter $G$. This
has the effect of making the
upper branch unstable. We now consider whether the instability of the
upper branch in the
limit $G \rightarrow 0$ is likely to be relevant during the initiation of
the BL, when one may expect the vortex sheet approximation to be valid.
Redimensionalizing Equation (\ref{uvp1}), we obtain
\ba
\label{gfreq}
\omega_1 \sim \Omega_K \eps^{1/2} i,
\ea
where $\Omega_K$ is the Keplerian velocity at the surface of the star.
During the initiation of the BL, we
expect disk material to be less dense than stellar material, which
means that $\eps \lesssim 1$. It then follows from Equation
(\ref{gfreq}) that the characteristic growth rate of the instability
is $\lesssim \Omega_K$ and is independent of
the wavenumber. 

If the BL is radially thin, as would be expected
during the initiation phase, then the growth rate given by Equation
(\ref{gfreq}) is small compared to the shear rate $S \sim \Omega_K
R_*/\delta_{BL} \gg \Omega_K$. Considering that $S$ is the characteristic
growth rate for shear instabilities, if there are other
mechanisms for instability with growth rates proportional to $S$, they
will quickly become dominant. 

In particular, we have demonstrated that the growth rate of the 
sonic instability of a finite width shear layer at high Mach 
number is proportional to $S$. The sonic instability is similar in
nature to the Papaloizou-Pringle instability in that both are global
instabilities and cannot be derived from a local analysis. There are
two destabilizing mechanisms for the sonic instability. The first
corresponds to emission of radiation and gives instability over a broad
range of wavenumbers. The second corresponds to over-reflection of
trapped modes and results in sharply peaked, disconnected regions of instability in
$k$-space. Because sonic instabilities
operate on a much faster
timescale than the gravity mechanism for a vortex sheet, this makes
them
an appealing candidate for the initial stages of boundary
layer formation, when one might expect large shears to be
present. 

We mention that \citet{AlexakisMiles} have 
considered the Miles instability \citep{Mileswave} in the context of
boundary layer
formation. Specifically, they invoked the
Miles instability to generate mixing of WD and stellar
material and explain the enrichment in heavy elements observed in nova
explosions. The Miles instability was proposed as a mechanism for
generating waves over
water at low wind speeds and operates through a
resonant interaction between the wind and the water wave. Due to this
interaction, a
component of the pressure perturbation is created that is in phase with the
velocity of the air-water interface, and much like pumping a swing, swings
up the interface to large amplitudes. However, the Miles instability
was initially introduced as a way of explaining the formation of waves
on water at weak wind speeds, for which the air-water interface is
stable to the KH instability. During boundary layer formation,
however, large shears are generated, so we are not in the weak wind
regime, and the Miles instability is likely to be swamped by sonic
instabilities.

There are two main astrophysical implications of our findings. First, we 
demonstrate that the initiation of the BL is likely to take only very 
short amount time after the first material from the disk arrives to the 
stellar surface. This is because the sonic instabilities that we explored 
in this paper have an extremely short growth rate, which is very weakly
dependent on the density contrast between the disk and stellar
material.\footnote{This statement is not true for the supersonic KH 
instability with gravity investigated in \S \ref{compress_grav}, 
see Equation (\ref{uvp1}).} Thus, mixing of the two fluids starts
almost immediately after they come into contact.

Second, given the efficiency with which the sonic instability operates, it is 
likely that it may play important role also in more developed phases of 
the BL evolution. As long as the BL possesses some effective "boundaries" 
(e.g. sharp changes in the velocity of density behavior) and the gas flow 
within it is supersonic the purely hydrodynamic sonic instabilities are 
going to operate in it potentially providing means for continuing mixing 
and angular momentum transport inside the layer. Future numerical
calculations capable of following the nonlinear development of sonic 
instabilities should be able to address this issue.

\acknowledgements

We are grateful to Jim Stone for useful discussions. 
The financial support for this work is provided by 
the Sloan Foundation and NASA grant NNX08AH87G.

\appendix

%%%%%%%%%%%%%%%%%%%%%%%%%%%%%%%%%%%%%%%%%%%

\section{Reduction of Equations}
\label{Reduction}

Defining $S \equiv \cp d\Omega/d \cp$ to be the shear rate, and assuming
$\delta_{BL} \ll R_*$ and $\Omega_* \ll \Omega_K$ we have that $2B \approx S \sim \Omega_K
R_*/\delta_{BL} \gg \Omega_K$. We next
note that $|\bo|^2 \ge (\Im[\bo])^2$ with exact equality at the location
of a critical layer, where $\Re[\bo] = 0$. According to \citet{Chimonas}, the
growth rate of the fastest growing mode in a
plane parallel stratified shear flow is of the order
\ba
\Im[\omega] \sim \frac{1}{4} S^2 - N^2,
\ea
where $N$ is the Brunt-V\"{a}is\"{a}l\"{a} frequency (Equation (\ref{eqBrunt})).
If we assume that shear instabilities provide the turbulence in the BL,
then we can estimate that $ |\bo|^2 \gtrsim S^2$. Using this estimate
on the left hand side of Equation (\ref{deltaueq}), we find that
the first term, $\bo^2$, dominates the third term, $\kappa^2 \approx 2\Omega S$  by a factor of $R_*/\delta_{BL} \gg 1$, so we can
ignore the third term. Next, we
compare the second to last term $\bo \gb/s^2$ to the last term $2
\Omega m / \cp$ on the right hand
side of Equation (\ref{deltaueq}). Defining $h_s \equiv s^2/\gb$,
dividing the second to last term by
the last term, and ignoring constants of order unity we have $ \bo R_* /h_s m \Omega
\gtrsim S R_* / h_s m \Omega \sim R_*^2 / h_s \delta_{BL} m$. Next, we
make the additional reasonable assumption that the fastest growing
mode has $m/R_* \sim h_s^{-1}$, since $h_s$ is of order the scale
height and sets the natural length scale in the problem. Continuing
our line of reasoning, we then have $R_*^2 / h_s \delta_{BL} m \sim
R_*/\delta_{BL} \gg 1$. This means that the last term in equation
(\ref{deltaueq}) is negligible compared to the second to last term and
can be ignored. Finally, we compare the first term, $C_L \bo$, and the
second term, $2Bm/\cp$, on the right hand side of Equation
(\ref{deltaPeq}). Dividing the first term by the second term, using
$2B \approx S$, $|\bo|^2 \gtrsim S^2$ and
$m/R_* \sim h_s^{-1}$, and noting that $C_L \sim h_s^{-1}$, we have $C_L
\bo R_*/m S \gtrsim h_s S/ h_s S \sim 1$. Thus, both terms can
potentially be of comparable magnitude and both must be retained.

Keeping only the dominant terms for $\delta_{BL}
\ll R_*$, using $\cp \approx R_*$, and assuming that $\rho'/\rho \gg
R_*^{-1}$ and $\delta u'/\delta u \gg R_*^{-1}$, Equations 
(\ref{deltaPeq}) and (\ref{deltaueq}) finally reduce to Equations 
(\ref{deltaPeq1}) and (\ref{deltaueq1}).

%%%%%%%%%%%%%%%%%%%%%%%%%%%%%%%%%%%%%%%%%%%

\section{Generalized Rayleigh Equation and Boundary Conditions}

%%%%%%%%%%%%%%%%%%%%%%%%%%%%%%%%%%%%%%%%%%%

\subsection{Derivation of the Generalized Rayleigh Equation}
\label{gRapp}

We start with Equations (\ref{deltaueq2}) and (\ref{deltaPeq2}). Using the
definitions from \S \ref{goveqn}, Equation
(\ref{deltaueq2}) can be written as
\ba
\label{vzifactoreq}
\delta u' + C_1(x) \delta u &=& C_2(x) \\
C_1(x) &=& -(k_g + W'/W) \\
C_2(x) &=& i(1-W^2/s^2) k_y \delta P/\rho W,
\ea
and similarly Equation (\ref{deltaPeq2}) can be written as
\ba
\label{pifactoreq}
\delta P' + C_3(x) \delta P &=& C_4(x) \\
C_3(x) &=& k_g \\
C_4(x) &=& -i \rho(k_y^2 W^2 + g (k_g + k_s)) \delta u/k_y W.
\ea
We can use the integrating factor method of solving first order
differential equations to make a change of variables and get rid of the
$\delta u$ term in Equation (\ref{vzifactoreq}) and the $\delta P$
term in Equation (\ref{pifactoreq}). Making these changes of variable, we have
\ba
\label{dqeq}
\delta q' &=& \frac{(1-W^2/s^2) k_y \delta P}{\rho W^2 f}, \ \delta q \equiv (i
W f)^{-1} \delta u\\
\label{djeq}
\delta \tilde{P}' &=& -\frac{i f \rho(k_y^2 W^2 + g(k_g + k_s))\delta
  u}{k_y W}, \ \delta \tilde{P} \equiv f \delta P.
\ea
We can combine Equations (\ref{dqeq}) and (\ref{djeq}) into a single
  equation for $\delta q$:
\ba
\label{2orderqeq}
\left(\frac{\tilde{\rho} W^2 \delta q'}{1-W^2/s^2}\right)' = \tilde{\rho}(k_y^2W^2 +
  g(k_g + k_s)) \delta q.
\ea
This is the same basic form as \citet{Chimonas}. Since Equation
  (\ref{2orderqeq}) is a second order differential equation for
  $\delta q$, we
  can convert it to standard form (get rid of the first order term)
  with the change of variable  
  \ba 
  \label{dphieq}
  \delta \phi = \tilde{W} \delta q /k_y.
  \ea
  Making this change of variable
  yields Equation (\ref{gen_Rayleigh}) which is the generalized
  Rayleigh equation \citep{Alexakis}.

%%%%%%%%%%%%%%%%%%%%%%%%%%%%%%%%%%%%%%%%%%%%%

\subsection{Derivation of the Boundary Conditions at the Interface}
\label{BCapp}

We present here a physically motivated derivation of the interfacial boundary
conditions, Equations (\ref{BC_contact}) and (\ref{BC_derivative}). The two classical boundary conditions
for the vortex sheet that should be satisfied at the
interface are (1) that the two fluids
should stay in contact at the interface, and (2) that the
force exerted on the lower fluid by the upper fluid is equal and
opposite to the force exerted on the upper fluid by the lower
fluid. These can be expressed as
\ba
\label{xibc}
\delta \xi_+ &=& \delta \xi_- \\
\label{Pbc}
\Delta P_+ &=& \Delta P_-,
\ea
where $\delta \xi$ is the displacement of the interface in the $x$-direction, and
$\Delta$ denotes the Lagrangian
differential. However, we must formulate both of these boundary conditions in
terms of the generalized streamfunction perturbation $\delta \phi$.

We begin with the condition $\delta \xi_+ = \delta \xi_-$. We start with the
relation
\ba
\label{Dxidt}
\frac{D\delta \xi}{Dt} = -i(\omega - k_yV_y)\delta \xi = \delta u,
\ea
where $D/Dt$, denotes the Lagrangian derivative. From Equations (\ref{xibc}) and (\ref{Dxidt}) we have
\ba
\frac{\delta u_+}{\omega - k_y V_{y}} = \frac{\delta u_-}{\omega +
  k_y V_{y}}.
\ea
Using the definitions of $\delta \phi$, $\tilde{W}$, and $\tilde{\rho}$ given
in \S \ref{goveqn}, and noting that $\tilde{\rho} = \rho$ at the
interface ($x=0$), we immediately get Equation (\ref{BC_contact}):
\ba
\frac{\delta \phi_+}{\tilde{W}_+} = \frac{\delta \phi_-}{\tilde{W}_-}.
\ea

For the second boundary condition, we begin by writing
\ba
\Delta P = \delta P - g \rho \delta \xi.
\ea
Using Equation (\ref{dqeq}) we can substitute for $\delta P$ in terms of
$\delta q$, which gives at the interface
\ba
\frac{\tilde{W}_+^2}{k_y}\delta q_+' - g_+ \rho_+ \delta \xi_+ =
\frac{\tilde{W}_-^2}{k_y}\delta q_-' - g_- \rho_- \delta \xi_-.
\ea
Using Equations (\ref{Dxidt}), (\ref{dqeq}), and (\ref{dphieq}) to substitute for
$\delta \xi$, $\delta q$, and $\delta u$ in terms of $\delta \phi$, we arrive at Equation (\ref{BC_derivative}):
\ba
\tilde{W}_+\delta \phi_+' - \tilde{W}_+' \delta \phi_+ - \frac{g_+ \tilde{\rho}_+
  \delta \phi_+}{\tilde{W}_+} = \tilde{W}_- \delta \phi_-' -
\tilde{W}_-' \delta \phi_- - \frac{g_- \tilde{\rho}_-
  \delta \phi_-}{\tilde{W}_-}.
\ea

\section{Finite Width Shear Layer}
\label{Glatzelapp}

We derive here the dispersion relation for a finite width shear layer
in the absence of gravity and for a constant shear. Our analysis
extends the work of \citet{Glatzel} to arbitrary density ratios above
and below the shear layer. We will use his notation in this section
and consider the pressure
perturbation $\delta P$ rather than the generalized stream function
$\delta \phi$. Where appropriate, we describe how to transform the
results back into the notation used in the body of the text. 

%%%%%%%%%%%%%%%%%%%%%%%%%%%%%%%%%%%%%%%%

\subsection{Setup of the Problem}

Consider the velocity profile
\ba
V(x) = \left\{
     \begin{array}{lr}
       -1, \ x < -1 \\
       x, \ -1 \le x \le 1 \\
       1, \ x > 1,
     \end{array}
   \right.
\ea
and the density profile
\ba
\rho(x) = \left\{
     \begin{array}{lr}
       \rho_-, \ x < -1 \\
       \rho_0, \ -1 \le x \le 1 \\
       \rho_+ , \ x > 1.
     \end{array}
   \right.
\ea
The adiabatic index and equilibrium pressure are everywhere constant
so we have $\rho_-s_-^2 = \rho_+s_+^2 = \rho_0s_0^2$.
The perturbations are assumed to be of the form $\delta P  = \delta
f(x) \exp[i (k_y y - \omega t)]$. We define $M = 1/s_0$ to be
the Mach number at $x=1$ inside the shear layer. We also define
\ba
\epsilon_- &=& \rho_0/\rho_-  \\
\epsilon_+ &=& \rho_0/\rho_+
\ea
and adopt the following quantities from \citet{Glatzel}
\ba 
\label{sigma_eq}
\bs &\equiv& W = -\frac{\bo}{k_y}\\
\label{Q_eq}
Q &\equiv& \bs^{-1/2} \delta P \\
\label{zeta_eq}
\zeta &=& i k_y M \bs^2 \\
\label{chi_eq}
\chi &=& \frac{i}{4}\frac{k_y}{M} \\
\mu &=& \frac{3}{4}.
\ea 

\citet{Glatzel} has shown that $Q$ satisfies Whittaker's
equation
\ba
\frac{d^2 Q}{d \zeta^2} + \left(-\frac{1}{4} + \frac{\chi}{\zeta} +
\frac{1/4 - \mu^2}{\zeta^2}\right)Q = 0
\ea
and inside the shear layer has the solution
\ba
\label{whit_sol}
Q = c_1 M_{\chi, \mu}(\zeta) + c_2 M_{\chi,-\mu}(\zeta),
\ea
where $c_1$ and $c_2$ are constants and $M_{\chi, \mu}(\zeta)$ is a
Whittaker function. Outside the shear layer the
velocity is constant, and the perturbation can be written as 
\ba
\label{Qeq}
Q \propto \left\{
     \begin{array}{lr}
       \exp\left[\pm k_y(1-\eps_-^{-1}M^2\bs^2)^{1/2}x \right], \ x < -1 \\
       \exp\left[\pm k_y(1-\eps_+^{-1}M^2\bs^2)^{1/2}x \right], \ x > 1
     \end{array}
   \right.
\ea
The signs in the
exponentials should be chosen based on the condition of outgoing
waves (\S \ref{ratiomach}).

%%%%%%%%%%%%%%%%%%%%%%%%%%%%%%%%%%%%

\subsection{Dispersion Relation}

To derive the dispersion relation, we must apply the contact
and pressure continuity boundary conditions at $x = \pm 1$ (Appendix
\ref{BCapp}). Since the velocity and hence $\bs$ are everywhere
continuous, the pressure continuity boundary conditions at $x = \pm 1$
are simply
\ba
\label{Qcond}
Q_\text{in} = Q_\text{out}
\ea 
where the subscripts ``in'' and ``out'' denote evaluation of a
quantity inside or outside the shear layer, respectively. Using 
the expression \citep{Glatzel}
\ba
\delta u = \frac{i}{k \rho \bs} \delta P',
\ea
the contact boundary conditions at $x = \pm 1$ can be written as
\ba
\delta u_\text{in} &=& \delta u_\text{out} \\
\rho_\text{out}\delta P_\text{in}' &=& \rho_\text{in} \delta
P_\text{out}' \\
\label{disp_eps}
\frac{1}{4 \zeta} + \frac{1}{Q_\text{in}}\frac{d Q_\text{in}}{d \zeta} &=& \pm
\frac{\rho_\text{in}/\rho_\text{out}}{2}\left(\frac{\rho_\text{out}}{\rho_\text{in}}- 4
\frac{\chi}{\zeta}\right)^{1/2}.
\ea
In deriving the expression (\ref{disp_eps}), we have made use of
relations (\ref{sigma_eq}-\ref{chi_eq}), and also the results
(\ref{Qeq}) and (\ref{Qcond}). The sign in Equation (\ref{disp_eps}) should
be chosen on the basis of outgoing waves, and the lower sign should be chosen at $x=1$ and the
upper sign at $x=-1$. We also point out that
$\rho_\text{in}/\rho_\text{out} = \eps_\pm$ at $x=\pm 1$.

Next, we substitute Equation (\ref{whit_sol}) into Equation
(\ref{disp_eps}) at $x=\pm 1$, and set $c_1 = 1$, which sets the
normalization. Using the property of the Whittaker function
\citep{AbramowitzStegun} that 
\ba
\zeta \frac{d M_{\chi,\mu}}{d \zeta} = \left(\frac{\zeta}{2} -
\chi \right) M_{\chi,\mu} + \left(\frac{1}{2} + \chi + \mu
\right) M_{\chi + 1, \mu},
\ea
we can derive the dispersion relation
\begin{multline}
\label{dis_finite}
\frac{\left[1-4\chi + 2 \zeta_+ + 2 \eps_+ \zeta_+ \left(\eps_+^{-1}-4\frac{
      \chi}{\zeta_+}\right)^{1/2}\right] M_{\chi,\frac{3}{4}}(\zeta_+) + (4\chi +
  5)M_{\chi+1,\frac{3}{4}}(\zeta_+)}{\left[1- 4\chi + 2\zeta_+ +
      2 \eps_+ \zeta_+\left(\eps_+^{-1} - 4 \frac{
      \chi}{\zeta_+}\right)^{1/2} \right] M_{\chi,-\frac{3}{4}}(\zeta_+) + (4\chi
  -1)M_{\chi+1,-\frac{3}{4}}(\zeta_+)} = \\
  \frac{\left[1-4\chi + 2 \zeta_- - 2 \eps_- \zeta_- \left(\eps_-^{-1}
      -4 \frac{
      \chi}{\zeta_-}\right)^{1/2}\right] M_{\chi,\frac{3}{4}}(\zeta_-) + (4\chi +
  5)M_{\chi+1,\frac{3}{4}}(\zeta_-)}{\left[1- 4\chi + 2\zeta_- -
      2 \eps_- \zeta_-\left(\eps_-^{-1} - 4 \frac{
      \chi}{\zeta_-}\right)^{1/2} \right] M_{\chi,-\frac{3}{4}}(\zeta_-) + (4\chi
  -1)M_{\chi+1,-\frac{3}{4}}(\zeta_-)}.
\end{multline}

Using the relation between the confluent hypergeometric function and
the Whittaker function \citep{Glatzel}, Equation (\ref{dis_finite}) can be written
in terms of the confluent hypergeometric function as
\begin{multline}
\label{dis_hyp}
\bs_+^3 \frac{\left[1-4\chi + 2 \zeta_+ + 2 \eps_+ \zeta_+
      \left(\eps_+^{-1} - 4 \frac{
      \chi}{\zeta_+}\right)^{1/2}\right] \HF(\frac{5}{4}-\chi,\frac{5}{2},\zeta_+) + (4\chi +
  5)\HF(\frac{1}{4}-\chi,\frac{5}{2},\zeta_+)}{\left[1- 4\chi + 2\zeta_+ +
      2 \eps_+ \zeta_+\left(\eps_+^{-1} - 4 \frac{
      \chi}{\zeta_+}\right)^{1/2} \right] \HF(-\frac{1}{4}-\chi,-\frac{1}{2},\zeta_+) + (4\chi
  -1)\HF(-\frac{5}{4} - \chi,-\frac{1}{2},\zeta_+)} = \\
  \bs_-^3 \frac{\left[1-4\chi + 2 \zeta_- - 2 \eps_- \zeta_-
      \left(\eps_-^{-1} - 4\frac{
      \chi}{\zeta_-}\right)^{1/2}\right]
      \HF(\frac{5}{4}-\chi,\frac{5}{2},\zeta_-) + (4\chi +
  5)\HF(\frac{1}{4}-\chi,\frac{5}{2},\zeta_-)}{\left[1- 4\chi + 2\zeta_- -
      2 \eps_- \zeta_-\left(\eps_-^{-1} - 4 \frac{
      \chi}{\zeta_-}\right)^{1/2} \right]
      \HF(-\frac{1}{4}-\chi,-\frac{1}{2},\zeta_-) + (4\chi
  -1)\HF(-\frac{5}{4} - \chi,-\frac{1}{2},\zeta_-)}.
\end{multline}

Equation (\ref{dis_hyp}) is a generalization of the dispersion
relations considered by \citet{Glatzel}. For instance, taking
$\eps_\pm \rightarrow 0$ we recover his dispersion relation (5.23) for
the reflecting boundary condition 
$\delta u(x=\pm 1) = 0$, taking $\eps_\pm \rightarrow \infty$ we
recover his dispersion relation (5.24) for a vacuum boundary
condition $\delta P(x=\pm 1) = 0$, and
taking $\eps_\pm = 1$ we recover his dispersion relation (5.38)
  for an infinite fluid with uniform density everywhere.

%%%%%%%%%%%%%%%%%%%%%%%%%%%%%%%%%%%%%%%%

\subsection{The Vortex Sheet Limit}
\label{KH_conv}

We can show that Equation (\ref{dis_hyp}) reduces to the
dispersion relation for the vortex sheet, (Equation
(\ref{compressdis})), by taking the limit $k_y \rightarrow 0$. In this
limit, $\zeta \rightarrow 0$ and  $\chi \rightarrow 0$, but $\bs$, the
phase speed in the comoving frame, is finite, and
\ba
4\frac{\chi}{\zeta} = (M^2 \bs^2)^{-1}
\ea
is also finite. We also note the property of the hypergeometric function
that in the limit $\zeta \rightarrow 0$
\ba
\HF(l,m,\zeta) = 1 + \frac{l}{m} \zeta + \mathcal{O}(\zeta^2).
\ea
Using these relations, Equation (\ref{dis_hyp}) can be reduced to the
form 
\ba
\bs_+^2 \eps_-(M^2 \bs_-^2 \eps_-^{-1} - 1)^{1/2} = \bs_-^2 \eps_+
(M^2 \bs_+^2 \eps_+^{-1} - 1)^{1/2}
\ea
Equation (\ref{compressdis}) can then be obtained by setting $\eps_+ =
1$, $\eps_- = \eps$, and using the relation $M = 1/s_+$.

\end{document}